\documentclass[12pt]{iopart}
\pdfoutput=1
\usepackage{iopams}
\usepackage{amssymb, epsfig}
\usepackage{latexsym}

\usepackage{graphicx}
\usepackage{color}

\begin{document}

\def\debproof{\noindent {\bf Proof.} }
\def\finproof{\hfill {\small $\Box$} \\}
\renewcommand{\theequation}
{\arabic{section}.\arabic{equation}}

\title[RTM for Inverse Scattering Problems]{Reverse Time Migration for Extended Obstacles: Acoustic Waves}
\author{Junqing Chen$^1$, Zhiming Chen$^2$,
Guanghui Huang$^2$ }
\address{$^1$ Department of Mathematical
Sciences, Tsinghua University,
Beijing 100084, China}
\address{$^2$ LSEC, Institute of Computational Mathematics, Academy of
Mathematics and Systems Science, Chinese Academy of Sciences,
Beijing 100190, China}

\begin{abstract}
We consider the resolution of the single frequency reverse time migration (RTM) method for extended targets
without the assumption of the validation of geometric optics approximation. The resolution analysis, which applies in both penetrable and non-penetrable obstacles with sound soft or impedance boundary condition on the boundary of the obstacle, implies that the imaginary part of the cross-correlation imaging functional is always positive and thus may have better stability properties. Numerical experiments are included to illustrate the powerful imaging quality and to confirm our resolution results.
\end{abstract}
\maketitle

\newcommand{\RR}{\mathcal{R}}
\newtheorem{lem}{Lemma}[section]
\newtheorem{prop}{Proposition}[section]
\newtheorem{cor}{Corollary}[section]
\newtheorem{thm}{Theorem}[section]
\newtheorem{rem}{Remark}[section]
\newtheorem{alg}{Algorithm}[section]
\newtheorem{assum}{Assumption}[section]
\newtheorem{definition}{Definition}[section]

\newcommand{\bL}{\mathbf{L}}
\newcommand{\bH}{\mathbf{H}}
\newcommand{\bW}{\mathbf{W}}
\newcommand{\bP}{\mathbf{P}}
\newcommand{\bQ}{\mathbf{Q}}
\newcommand{\bp}{\mathbf{p}}
\newcommand{\bq}{\mathbf{q}}
\newcommand{\uL}{u_{_{\rm L}}}
\newcommand{\vL}{v_{_{\rm L}}}
\newcommand{\tuL}{\tilde u_{_{\rm L}}}
\newcommand{\tvL}{\tilde v_{_{\rm L}}}
\newcommand{\fL}{f_{_{\rm L}}}
\newcommand{\gL}{g_{_{\rm L}}}
\newcommand{\bpL}{\bp_{_{\rm L}}}
\newcommand{\bqL}{\bq_{_{\rm L}}}
\newcommand{\tbpL}{\tilde{\bp}_{_{\rm L}}}
\newcommand{\tbqL}{\tilde{\bq}_{_{\rm L}}}
\newcommand{\tbpLf}{\tilde{\bp}_{_{\rm L,1}}}
\newcommand{\tbpLs}{\tilde{\bp}_{_{\rm L,2}}}
\newcommand{\tbqLf}{\tilde{\bq}_{_{\rm L,1}}}
\newcommand{\tbqLs}{\tilde{\bq}_{_{\rm L,2}}}
\newcommand{\bn}{\nu}
\newcommand{\bv}{\mathbf{v}}
\newcommand{\om}{\omega}
\newcommand{\pa}{\partial}
\newcommand{\la}{\langle}
\newcommand{\ra}{\rangle}
\newcommand{\lla}{\la{\hskip -2pt}\la}
\newcommand{\rra}{\ra{\hskip -2pt}\ra}
\newcommand{\jj}{\|{\hskip -0.8pt} |}
\newcommand{\al}{\alpha}
\newcommand{\ze}{\zeta}
\newcommand{\si}{\sigma}
\newcommand{\ep}{\varepsilon}
\newcommand{\na}{\nabla}
\newcommand{\vp}{\varphi}
\newcommand{\ga}{\gamma}
\newcommand{\Ga}{\Gamma}
\newcommand{\Om}{\Omega}
\newcommand{\de}{\delta}
\newcommand{\Th}{\Theta}
\newcommand{\De}{\Delta}
\newcommand{\Lam}{\Lambda}
\newcommand{\lam}{\lambda}
\newcommand{\tri}{\triangle}
\newcommand{\lj}{[{\hskip -2pt} [}
\newcommand{\rj}{]{\hskip -2pt} ]}
\newcommand{\bks}{\backslash}
\newcommand{\diam}{\mathrm{diam}}
\newcommand{\osc}{\mathrm{osc}}
\newcommand{\meas}{\mathrm{meas}}
\newcommand{\dist}{\mathrm{dist}}

\newcommand{\mL}{\mathscr{L}}
\newcommand{\cT}{{\cal T}}
\newcommand{\cM}{{\cal M}}
\newcommand{\cE}{{\cal E}}
\newcommand{\cL}{{\cal L}}
\newcommand{\cF}{{\cal F}}
\newcommand{\cB}{{\cal B}}
\newcommand{\PML}{{\rm PML}}
\newcommand{\FEM}{{\rm FEM}}
\newcommand{\rd}{\,\mathrm{d}}

\renewcommand{\i}{\mathbf{i}}
\newcommand{\R}{{\mathbb{R}}}
\newcommand{\Z}{{\mathbb{Z}}}
\newcommand{\C}{{\mathbb{C}}}
\renewcommand{\Re}{\mathrm{Re}\,}
\renewcommand{\Im}{\mathrm{Im}\,}
\renewcommand{\div}{\mathrm{div}}
\newcommand{\curl}{\mathrm{curl}}
\newcommand{\Curl}{\mathbf{curl}}

\newcommand{\be}{\begin{eqnarray}}
\newcommand{\ee}{\end{eqnarray}}
\newcommand{\ben}{\begin{eqnarray*}}
\newcommand{\een}{\end{eqnarray*}}
\newcommand{\nn}{\nonumber}

\section{Introduction}\label{section1}

In this paper we propose and study an imaging algorithm to find the support of an unknown obstacle embedded in a known background medium
from a knowledge of scattered acoustic waves measured on a given acquisition surface which is assumed to be far away from the obstacle. The
algorithm does not require any a priori information of the physical properties of the obstacle such as penetrable
or non-penetrable, and for non-penetrable obstacles, the type of boundary conditions on the boundary of the obstacle.

Let the obstacle occupy a bounded Lipschitz domain $D\subset\R^2$ with $\nu$ the unit outer normal to its boundary $\Ga_D$.
We assume the incident wave is emitted by a point source located at $x_s$ on a closed surface $\Ga_s$ away from the obstacle.
For penetrable obstacles $D$, the measured wave $u$ is the solution of the following acoustic scattering problem:
\be
& &\De u+k^2n(x)u= -\de_{x_s}(x)\ \ \ \ \mbox{in }\R^2,\label{p1}\\
& &\sqrt{r}\left(\frac{\pa u}{\pa r}-\i ku\right)\to 0\ \ \ \ \mbox{as }r\to\infty,\ \ r=|x|,\label{p2}
\ee
where $k>0$ is the wave number, $n(x)\in L^\infty(D)$ is a positive scalar function supported in $D$, $\de_{x_s}$ is the Dirac source located at $x_s$. The condition (\ref{p2}) is the well-known Sommerfeld radiation condition. For non-penetrable obstacles $D$, the measured wave $u$
is the solution of the following scattering problem:
\be
& &\De u+k^2u= -\de_{x_s}(x)\ \ \ \ \mbox{in }\R^2,\label{p3}\\
& &u=0\ \ \mbox{or }\ \ \frac{\pa u}{\pa\nu}+\i k\eta(x)u=0\ \ \ \ \mbox{on }\Ga_D,\label{p4}\\
& &\sqrt{r}\left(\frac{\pa u}{\pa r}-\i ku\right)\to 0\ \ \ \ \mbox{as }r\to\infty,\ \ r=|x|,\label{p5}
\ee
where $\eta(x)\ge 0$ is a bounded function on $\Ga_D$. The Dirichlet boundary condition $u=0$ on $\Ga_D$ corresponds to the sound soft obstacle. The second condition on $\Ga_D$ in (\ref{p4}) is the impedance condition and it reduces to the sound hard obstacle when $\eta(x)=0$. The existence and uniqueness of the problem (\ref{p1})-(\ref{p2}) such that $u^s=u-u^i$ in $H^1_{\rm loc}(\R^2)$ and
the problem (\ref{p3})-(\ref{p5}) such that $u^s=u-u^i$ in $H^1_{\rm loc}(\R^2\backslash\bar D)$ is well-known \cite{colton-kress, mclean00, ccm01}, where $u^i(x)=\frac{\i}{4}H^{(1)}_0(k|x-x_s|)$ is the fundamental solution of the Helmholtz equation, $H^{(1)}_0(z)$ is the Hankel function of the first type and order zero.

The direct methods for solving inverse scattering problems have drawn considerable interest in the literature in recent years. One example is the
MUltiple SIgnal Classification (MUSIC) method which was first proposed in the signal processing in Schmidt \cite{music}.  It was used for imaging point scatterers under Born approximation for well-resolved targets by Devaney \cite{Devaney} and extended to locate small inclusions in Bruhl, Hanke, and Vogelius \cite{BHV} and Ammari \cite{anomaly}. The key ingredient in the MUSIC algorithm is to construct a basis function for the signal space via singular value decomposition (SVD) of the multi-static response matrix (MSR). For extended targets or cracks, the singular values of MSR matrix may decrease continuously without significant gap between signal space and null (noise) space.

The other class of direct methods for inverse scattering problems includes the linear sampling method of Colton  and Kirsch \cite{LSM}, the factorization method of Kirsch in \cite{kirsch_1998}, and the point source method of Potthast \cite{p96}. These methods are motivated by the uniqueness proof of inverse scattering problems and are based on the fact that the point source incident wave becomes singular as the source approaches to the boundary of scatterers (see Colton and Kress \cite{ck06}). We refer to Potthast \cite{potthast} and Kirsch and Grinberg \cite{fm_book} for more information on this class of methods.

The reverse time migration (RTM) or the closely related prestack depth migration methods are nowadays widely used in exploration geophysics (see e.g., Berkhout \cite{ber84}, Claerbout \cite{cla85}). It is originated in the simple setting of exploding reflector model. For imaging the complex medium in practical applications, the analysis of the migration method is usually based on the high frequency assumption so that the geometric optics approximation can be used (see e.g., Bleistein, Cohen and Stockwell \cite{bcs}). Our work is inspired by the recent study of RTM method in a noisy environment in Garnier \cite{g11} in which the Helmholtz-Kirchhoff identity is used to study the resolution of RTM for imaging small inclusions. Helmholtz-Kirchhoff identity plays an important role in the inverse source problem, see Bojarski \cite{boj}.

The purpose of this paper is to provide a new mathematical understanding of the RTM method for extended obstacles without the assumption of geometric optics approximation. We study the resolution of the RTM method for both penetrable and non-penetrable obstacles. As the outcome of our resolution analysis we propose to use the imaginary part of the cross-correlation RTM functional. We show that this new imaging functional enjoys the nice feature that it is always positive and thus may have better stability properties. We will extend the results in this paper to study the RTM method for imaging extend targets using  electromagnetic and elastic waves in forthcoming papers.

The rest of this paper is organized as follows. In section 2 we introduce the RTM algorithm. In section 3 we study the resolution of the imaging algorithm in section 2 for both the penetrable and non-penetrable obstacles. In section 4 we report extensive numerical experiments to
show the competitive performance of the new RTM algorithm.

\section{The reverse time migration method}\label{section2}

In this section we introduce the RTM method for inverse scattering problems. We assume that there are $N_s$ transducers uniformly distributed on $\Ga_s=\pa B_s$ and $N_r$ transducers uniformly distributed on $\Ga_r=\pa B_r$. $B_s$ and $B_r$ are the circles of  radius $R_s$ and $R_r$, respectively. We denote by $\Om$ the sampling domain in which the obstacle is sought. We assume the obstacle $D\subset\Om$ and $\Om$ is inside in $B_s$, $B_r$.

Let $G(x,y)$ be the fundamental solution of the Helmholtz equation
\ben
\De G(x,y)+k^2G(x,y)=-\de_y(x)\ \ \ \ \mbox{in }\R^2.
\een
Let $u^i(x,x_s)=G(x,x_s)$ be the incident wave and $u^s(x_r,x_s)=u(x_r,x_s)-u^i(x_r,x_s)$ be the scattered field measured at $x_r$, where $u(x,x_s)$ is the solution of the problem either (\ref{p1})-(\ref{p2}) or (\ref{p3})-(\ref{p5}). The matrix $(u^s(x_r,x_s))^{N_r\times N_s}_{i,j=1}$ is called the multi-static response matrix (MSR) in the literature.

Our RTM algorithm consists of two steps. The first step is the back-propagation in which we back-propagate the complex conjugated (time reversed) data $\overline{u(x_r,x_s)}$ into the domain. The second step is the cross-correlation in which we compute the imaginary part of the cross-correlation of the incident field and the back-propagated field.

\begin{alg} {\sc (Reverse time migration algorithm)} \\
Given the data $u^s(x_r,x_s)$ which is the measurement of the scattered field at $x_r$ when the source is emitted at $x_s$, $s=1,\dots, N_s$ and $r=1,\dots,N_r$. \\
$1^\circ$ Back-propagation: For $s=1,\dots,N_s$, compute the solution $v_b$ of the following problem:
\be
& &\Delta v_b(x,x_s)+k^2v_b(x,x_s)=\frac{|\Ga_r|}{N_r}\sum^{N_r}_{r=1}\overline{u^s(x_r,x_s)}\de_{x_r}(x)\ \ \ \ \mbox{in }\R^2,\label{b1}\\
& &\sqrt{r}\left(\frac{\pa v_b}{\pa r}-\i kv_b\right)\to 0\ \ \ \ \mbox{as }r\to\infty.\label{b2}
\ee
$2^\circ$ Cross-correlation: For $z\in\Om$, compute
\be\label{cor1}
I(z)=k^2\cdot\Im\left\{\frac{|\Ga_s|}{N_s}\sum^{N_s}_{s=1}u^i(z,x_s)v_b(z,x_s)\right\}.
\ee
\end{alg}

Taking the imaginary part of the correlation of the incidence field and the back-propagated field is motivated by the resolution analysis in the next section where we show that the imaginary part of the correlation functional is a positive function and thus is more stable than the real part of the correlation functional. By using the fundamental solution we can represent the solution $v_b$ of (\ref{b1})-(\ref{b2}) as
\ben
v_b(z,x_s)=-\frac{|\Ga_r|}{N_r}\sum^{N_r}_{r=1}G(z,x_r)\overline{u^s(x_r,x_s)},
\een
which implies
\be\label{cor2}
 \fl \qquad I(z)=-k^2\cdot\Im\left\{\frac{|\Ga_s||\Ga_r|}{N_sN_r}\sum^{N_s}_{s=1}\sum^{N_r}_{r=1}G(z,x_s)G(z,x_r)\overline{u^s(x_r,x_s)}\right\}\ \ \ \ \forall z\in\Om.
\ee
This formula is used in all our numerical experiments in section 4.

Noticing that for $z\in\Om$ which is a subdomain of $\Om_s$, $G(z,x_s)$ is a smooth function in $x_s\in\Ga_s$. Similarly, $G(z,x_r)$ is smooth in
$x_r\in\Ga_r$. We also know that since $u^s=u-u^i$ is the scattering solution of (\ref{p1})-(\ref{p2}) or (\ref{p3})-(\ref{p4}),
$u^s(x_r,x_s)$ is also smooth in
$x_r,x_s$. Therefore, the imaging functional $I(z)$ in (\ref{cor2}) is a good trapezoid quadrature approximation of the following continuous functional:
\be\label{cor3}
\fl \qquad \hat I(z)=-k^2\cdot\Im\int_{\Ga_r}\int_{\Ga_s}G(z,x_s)G(z,x_r)\overline{u^s(x_r,x_s)}ds(x_s)ds(x_r)\ \ \ \ \forall z\in\Om.
\ee
This formula is the starting point of our resolution analysis in the next section.

\section{The resolution analysis}\label{section3}

In this section we study the resolution of the imaging functional in (\ref{cor3}). We start by recalling the the Helmholtz-Kirchhoff identity \cite{boj}.

\begin{lem}\label{lem:3.1} Let $\mathcal{D}$ be a bounded Lipschitz domain in $\R^2$ with $\bn$ being the unit outer normal to its boundary; we have
\begin{equation*}
 \fl  \qquad  \int_{\partial\mathcal{D}} \left (\overline{G(x,\xi)}\frac{\partial G(\xi,y)}{\pa\bn}-\frac{\partial\overline{G(x,\xi)}}{\pa\bn}G(\xi,y)\right )ds(\xi)  =2\i\,{\rm{Im}}G(x,y)  \quad  \forall x,y \in\mathcal{D}.
\end{equation*}
\end{lem}

\debproof For the sake of completeness, we include a proof here. For any fixed $x\in \mathcal{D}$, since $\Im G(x,y)$ satisfies the Helmholtz equation, we obtain by the integral representation formula that for any $y\in\mathcal{D}$,
\ben
\Im G(x,y)=\int_{\pa\mathcal{D}}\left(\frac{\pa\Im G(x,\xi)}{\pa\nu}G(\xi,y)-\frac{\pa G(\xi,y)}{\pa\nu}\Im G(x,\xi)\right)ds(\xi).
\een
By $\Im G(x,\xi)=\frac 1{2\i}(G(x,\xi)-\overline{G(x,\xi)})$ we know the lemma follows if one can prove the following identity:
\be\label{hk0}
\int_{\pa D}\left(\frac{\pa G(\xi,y)}{\pa\nu}G(x,\xi)-\frac{\pa G(x,\xi)}{\pa\nu}G(\xi,y)\right)ds(\xi)=0\ \ \ \ \forall x,y\in\mathcal{D}.
\ee
Denote by $B_R$ a circle of radius $R>0$ that includes $\mathcal{D}$. Since $x,y\in\mathcal D$, $G(x,\cdot)$ and $G(\cdot,y)$ satisfy Helmholtz equation in $B_R\backslash\bar{\mathcal{D}}$. By integration by parts, we obtain
\ben
\hspace{-2cm}& &\int_{\pa D}\left(\frac{\pa G(\xi,y)}{\pa\nu}G(x,\xi)-\frac{\pa G(x,\xi)}{\pa\nu}G(\xi,y)\right)ds(\xi)\\
\hspace{-2cm}&=&\int_{\pa B_R}\left(\frac{\pa G(\xi,y)}{\pa\nu}G(x,\xi)-\frac{\pa G(x,\xi)}{\pa\nu}G(\xi,y)\right)ds(\xi)\\
\hspace{-2cm}&=&\int_{\pa B_R}\left[\left(\frac{\pa G(\xi,y)}{\pa\nu}-\i kG(\xi,y)\right)G(x,\xi)-\left(\frac{\pa G(x,\xi)}{\pa\nu}-\i kG(x,\xi)\right)G(\xi,y)\right]ds(\xi).
\een
This shows (\ref{hk0}) by letting $R\to\infty$ since $G(x,\xi)=O(|\xi|^{-1/2})$ and $\frac{\pa G(x,\xi)}{\pa\nu}-\i kG(x,\xi)=O(|\xi|^{-3/2})$ as $|\xi|\to\infty$. This completes the proof.
\finproof

A direct consequence of the Helmholtz-Kirchhoff identity is the following lemma which plays a key role in our resolution analysis, see also \cite{g11}.

\begin{lem}\label{lem:3.2}
We have
\be
& &k\int_{\Ga_s}\overline{G(x,x_{s})}G(x_{s},z)ds(x_s)=\Im G(x,z)+w_s(x,z)\ \ \ \ \forall x,z\in\Om, \label{hk1}\\
& &k\int_{\Ga_r}\overline{G(x,x_{r})}G(x_{r},z)ds(x_r)=\Im G(x,z)+w_r(x,z)\ \ \ \ \forall x,z\in\Om, \label{hk2}
\ee
where $|w_s(x,z)|+|\na_x w_s(x,z)|\le CR^{-1}_s, |w_r(x,z)|+|\na w_r(x,z)|\le CR^{-1}_r$ uniformly for any $x,z\in\Om$.
\end{lem}

\debproof The estimate for $|w_s(x,z)|$ in (\ref{hk1}) is a direct consequence of Lemma \ref{lem:3.1} and the following well-known asymptotic relations:
\ben
\hspace{-1cm}G(x,x_s)=O(R_s^{-1/2}),\ \ \frac{\pa G(x,x_s)}{\pa\bn}-\i k G(x,x_s)=O(R_s^{-3/2})\ \ \forall x\in \Om,x_s\in\Ga_s.
\een
The estimate for $|\na_x w_s(x,z)|$ follows again from Lemma \ref{lem:3.1} by using the following asymptotic relations:
\ben
\hspace{-1cm}\frac{\pa G(x,x_s)}{\pa x_j}=O(R_s^{-1/2}),\ \ \frac{\pa }{\pa x_j}\left(\frac{\pa G(x,x_s)}{\pa\bn}-\i k G(x,x_s)\right)=O(R_s^{-3/2}),
\een
for any $x\in \Om,x_s\in\Ga_s$, $j=1,2$.
Equation (\ref{hk2}) can be proved similarly. This completes the proof.
\finproof

To proceed we recall the definition of the Dirichlet-to-Neumann mapping $\Lambda_D:H^{1/2}(\partial D)\rightarrow H^{-1/2}(\partial D)$. For any $g\in H^{1/2}(\Ga_D)$, $\Lambda_D (g)=\frac{\pa w}{\pa\nu}\Big|_{\Ga_D}$, where $w\in H^1_{\rm loc}(\R^2\backslash\bar D)$ is the solution of the following scattering problem:
\be
& &\De w+k^2 w=0\ \ \ \ \mbox{on }\R^2\backslash\bar D,\label{w1}\\
& &w=g\ \ \mbox{on }\Ga_D,\ \ \ \ \sqrt{r}\left(\frac{\pa w}{\pa r}-\i kw\right)\to 0\ \ \mbox{as }r\to\infty.\label{w2}
\ee
The far field pattern $w_\infty(\hat x)$ of the solution $w$ to the scattering problem (\ref{w1})-(\ref{w2}) is defined by the asymptotic behavior
\be\label{far}
w(x)=\frac{e^{\i k|x|}}{|x|^{1/2}}\left\{w_\infty(\hat x)+O\left(\frac 1{|x|}\right)\right\},\ \ \ \ |x|\to\infty,
\ee
where $\hat x=x/|x|\in S^1:=\{x\in\R^2: |x|=1\}$.

\begin{lem}\label{lem:3.3}
Let $g\in H^{1/2}(\Ga_D)$ and $w$ be the radiation solution satisfying (\ref{w1})-(\ref{w2}); then
\begin{equation*}
-{\rm{Im}}\langle g,\Lambda_D(g)\rangle_{\Ga_D}=k\int_{S^1}|w_\infty(\hat x)|^2d\hat x > 0,
\end{equation*}
where $\la\cdot,\cdot\ra_{\Ga_D}$ represents the duality pairing between $H^{1/2}(\Gamma_D)$ and $H^{-1/2}(\Gamma_D)$.
\end{lem}

\debproof For the sake of completeness, we sketch a proof here. Let $B_R$ be a circle of radius $R$ that includes $D$. By integrating by parts one easily obtains
\ben
\la g,\Lam(g)\ra_{\Ga_D}=\int_{\Ga_D}w\frac{\pa\bar w}{\pa\nu}ds=-\int_{B_R\backslash\bar D}(|\na w|^2-k^2|w|^2)dx+\int_{\Ga_R}w\frac{\pa \bar w}{\pa r}ds.
\een
Thus, by using the Sommerfeld radiation condition, we have
\ben
\Im\la g,\Lam(g)\ra_{\Ga_D}=\lim_{R\to\infty}\Im\int_{\Ga_R}w\frac{\pa \bar w}{\pa r}ds=-k\lim_{R\to\infty}\int_{\Ga_R}|w|^2ds.
\een
This completes the proof by using (\ref{far}).
\finproof

The following stability estimate of the forward scattering problem problem will be used in our resolution analysis. The proof can be found in Zhang \cite{zhang94}
by using the method of limiting absorption principle where a general transmission problem in two locally perturbed half-spaces is studied.

\begin{lem}\label{lem:3.4}
Let $n \in L^{\infty}(D)$ be a positive scalar function supported in $D$ and the source $f\in L^2(\R^2)$ have compact support. Then the scattering problem
\ben
& &\Delta U+k^2n(x) U=f(x)\ \ \mbox{in }\R^2,\\
& &\sqrt{r}\left(\frac{\pa U}{\pa r}-\i kU\right)\to 0\ \ \ \ \mbox{as }r\to\infty,\ \ r=|x|,
\een
admits a unique solution $U\in H^{1}_{\rm loc}(\R^2)$. Moreover, there exists a constant $C>0$ such that $\|U\|_{H^1(D)}\le C \|f\|_{L^{2}(\R^2)}$.
\end{lem}

We remark that the constant $C$ in the lemma depends on the scatterer $D$ and the wave number $k$. The estimate of the explicit dependence of $C$ on the wave number $k$ when
$k$ is large draws considerable interests in the literature. We refer to the recent work of Chandler-Wilde and Monk \cite{cp08} and the references therein for
results in that direction. In this paper we are interested in the case when the wavelength $\lambda=2\pi/k$ is of comparable size of the scatterer, that is, the case when $k$ is not very large.

Now we are in the position to show the resolution theorem for the RTM algorithm in this paper. We first consider the case of penetrable obstacles.

\begin{thm}\label{thm:3.1} For any $z\in\Om$, let $\psi(x,z)$ be the radiation solution of the Helmholtz equation with penetrable scatterer $D$:
\be
& &\Delta\psi+k^2n(x)\psi=-k^2(n(x)-1)\Im G(x,z)\ \ \ \ \mbox{in }\R^2.\label{ps0}
\ee
Then if the measured field $u^s=u-u^i$ with $u$ satisfying the problem (\ref{p1})-(\ref{p2}), we have
\ben
\hat I(z)=k\int_{S^1}|\psi_\infty(\hat x,z)|^2d\hat x+w_{\hat I}(z)\ \ \ \ \forall z\in\Om,
\een
where $\|w_{\hat I}\|_{L^\infty(\Om)}\le C(R^{-1}_s+R^{-1}_r)$.
\end{thm}

\debproof Since $G(x,x_s)$ satisfies $\De G(x,x_s)+k^2G(x,x_s)=-\de_{x_s}(x)$ in $\R^2$, we know that $u^s=u-u^i$ satisfies
\ben
\De u^s(x,x_s)+k^2u^s(x,x_s)=k^2(1-n(x))u(x,x_s),
\een
which implies $u^s$ satisfies the Lippmann-Schwinger equation
\ben
u^s(x,x_s)=\int_{D}k^2(n(\xi)-1)u(\xi,x_s)G(x,\xi)d\xi.
\een
Then, by Lemma \ref{lem:3.2},
\ben
v_b(z,x_s)&=&\int_{\Ga_r}G(z,x_r)\overline{u^s(x_r,x_s)}ds(x_r)\\
&=&\int_{\Ga_r}\int_Dk^2(n(\xi)-1)\overline{u(\xi,x_s)}\,\overline{G(x_r,\xi)}G(z,x_r)d\xi ds(x_r)\\
&=&\int_Dk^2(n(\xi)-1)\left[\frac 1k\big(\Im G(\xi,z)+w_r(\xi,z)\big)\right]\overline{u(\xi,x_s)}d\xi,
\een
which yields after using (\ref{cor3}) that
\be\label{i1}
\hspace{-1cm}\hat I(z)&=&-k^2\Im\int_{\Ga_s}G(z,x_s)v_b(z,x_s)ds(x_s)\nn\\
&=&-k\ \Im\int_Dk^2(n(\xi)-1)\left[\frac 1k\big(\Im G(\xi,z)+w_r(\xi,z)\big)\right]v(\xi,z)d\xi,
\ee
where $v(\xi,z)=k\int_{\Ga_s}G(z,x_s)\overline{u(\xi,x_s)}ds(x_s)$. Since by the Lippmann-Schwinger equation
\ben
u(\xi,x_s)=G(\xi,x_s)+\int_Dk^2(n(y)-1)u(y,x_s)G(\xi,y)dy,
\een
we obtain by using Lemma \ref{lem:3.2} that
\ben
v(\xi,z)=(\Im G(\xi,z)+w_s(\xi,z))+\int_Dk^2(n(y)-1)v(y,z)\overline{G(\xi,y)}dy.
\een
Let $w(\xi,z)=v(\xi,z)-(\Im G(\xi,z)+w_s(\xi,z))$; then
\ben
w(\xi,z)=\int_Dk^2(n(y)-1)\left[w(y,z)+\big(\Im G(y,z)+w_s(y,z)\big)\right]\overline{G(\xi,y)}dy,
\een
and, consequently,
\ben
\overline{w(\xi,z)}=\int_Dk^2(n(y)-1)\left[\overline{w(y,z)}+ \big(\Im G(y,z)+\overline{w_s(y,z)}\big)\right]G(\xi,y)dy.
\een
This implies that $\overline{w(\xi,z)}$ is the radiation solution of the following Helmholtz equation
\ben
\De_\xi\overline{w(\xi,z)}+k^2\overline{w(\xi,z)}=-k^2(n(\xi)-1)\left[\overline{w(\xi,z)}+\big(\Im G(\xi,z)+\overline{w_s(\xi,z)}\big)\right],
\een
which is equivalent to
\ben
\De_\xi\overline{w(\xi,z)}+k^2n(\xi)\overline{w(\xi,z)}=-k^2(n(\xi)-1)\big(\Im G(\xi,z)+\overline{w_s(\xi,z)}\big).
\een
Now by (\ref{ps0}) we know that $\zeta(\xi,z)=\overline{w(\xi,z)}-\psi(\xi,z)$ satisfies
\ben
\De_\xi\zeta(\xi,z)+k^2n(\xi)\zeta(\xi,z)=-k^2(n(\xi)-1)\overline{w_s(\xi,z)}\ \ \ \ \mbox{in }\R^2,
\een
and the Sommerfeld radiation condition. By Lemma \ref{lem:3.4} and Lemma \ref{lem:3.2} we have
\be\label{zeta}
\|\zeta(\cdot,z)\|_{H^1(D)}\le Ck^2 \| (n(\cdot)-1)\|_{L^{\infty}(D)}\|\overline{w_s(\cdot,z)}\|_{L^2(D)}\le CR_s^{-1},
\ee
uniformly for $z\in\Om$. This implies by using Lemma \ref{lem:3.2} again that
\ben
v(\xi,z)&=&w(\xi,z)+\big(\Im G(\xi,z)+w_s(\xi,z)\big)\\
&=&\overline{\psi(\xi,z)}+\overline{\zeta(\xi,z)}+\big(\Im G(\xi,z)+w_s(\xi,z)\big),
\een
where $\|\zeta(\cdot,z)\|_{H^1(D)}+\|w_s(\cdot,z)\|_{L^2(D)}\le CR^{-1}_s$.
Now by (\ref{i1}) we obtain
\ben
\hat I(z)&=&-\Im\int_Dk^2(n(\xi)-1)\big(\Im G(\xi,z)+w_r(\xi,z)\big) v(\xi,z)d\xi\\
&=&-\ \Im\int_Dk^2(n(\xi)-1)\Im G(\xi,z)\overline{\psi(\xi,z)}d\xi+O(R^{-1}_s+R^{-1}_r).
\een
Now by the equation satisfied by $\psi(\xi,z)$ and integrating by parts we obtain
\ben
\hat I(z)&=&\Im\int_D\left[\De_\xi\psi(\xi,z)+k^2n(\xi)\psi(\xi,z)\right]\overline{\psi(\xi,z)}d\xi+O(R^{-1}_s+R^{-1}_r)\\
&=&\Im\int_{\Ga_D}\frac{\pa\psi(\xi,z)}{\pa\nu}\overline{\psi(\xi,z)}ds(\xi)+O(R^{-1}_s+R^{-1}_r)\\
&=&-\ \Im\la\psi(\cdot,z),\Lam_D(\psi(\cdot,z)\ra_D+O(R^{-1}_s+R^{-1}_r).
\een
This completes the proof by using Lemma \ref{lem:3.3}.
\finproof

We remark that since
$$\big(\De+k^2n(x) \big)\Im G(x,z)=k^2(n(x)-1)\Im G(x,z),$$
$\psi(x,z)$ can be viewed as the scattering solution of the Helmholtz equation with the incident wave $\Im G(x,z)$.  It is known that
$\Im G(x,z)=\Im \frac{\i}{4}H^{(1)}_0(k|x-z|)=\frac 14J_0(k|x-z|)$ which peaks when $x=z$ and decays as $|x-z|$ becomes large. Noticing that the source
in (3.17) is supported in $D$ because of $n(x)=1$ outside $D$. Thus the source in (\ref{ps0}) becomes small when $z$ moves away from $\pa D$ outside the scatterer. On the other hand, the source in (\ref{ps0}) will not be small if $z$ is inside $D$. Therefore we expect that the imaging functional will have contrast at the boundary of the scatterer $D$ and decays outside the scatterer. This is indeed confirmed in our numerical experiments.

Now we consider the resolution of the imaging functional in the case of non-penetrable obstacles. We only prove the results for the case of impedance boundary condition. The case of Dirichlet boundary condition is similar and simpler and is left to the interested readers. We need the following result about the forward
impedance scattering problem whose proof is similar to that for partially coated perfect conductor considered in \cite{ccm01}. It can also be proved by using
the method of limiting absorption principle, see e.g., Leis \cite[Remark 4.39]{leis}.

\begin{lem}\label{lem:3.5}
Let $\eta\ge 0$ be bounded and $g \in H^{-1/2}(\partial D)$. Then the scattering problem
\ben
& &\Delta U+k^2U=0\ \ \mbox{in }\R^2\backslash\bar D,\\
& &\frac{\pa U}{\pa\nu}+\i k\eta(x)U=-g\ \ \mbox{on }\Ga_D,\ \ \ \ \sqrt{r}\left(\frac{\pa U}{\pa r}-\i k U\right)\to 0\ \ \mbox{as }r\to\infty,
\een
admits a unique solution $U \in H^{1}_{\rm loc}(\R^2\backslash\bar D)$. Moreover, there exists a constant $C$ such that $\|U\|_{H^{1/2}(\Ga_D)}\le C\|g\|_{H^{-1/2}(\Ga_D)}$.
\end{lem}

\begin{thm}\label{thm:3.2}
For any $z\in\Om$, let $\psi(x,z)$ be the radiation solution of the Helmholtz equation
\be\label{ps1}
& &\De\psi(x,z)+k^2\psi(x,z)=0\ \ \ \ \mbox{in }\R^2\bks\bar D,\\
& &\frac{\pa\psi(x,z)}{\pa\nu}+\i k\eta(x)\psi(x,z)=-\left[\frac{\pa}{\pa\nu}+\i k\eta(x)\right]\Im G(x,z)\ \ \ \ \mbox{on }\Ga_D.\label{ps2}
\ee
Then if the measured field $u^s=u-u^i$ with $u$ satisfying the problem (\ref{p3})-(\ref{p5}) with the impedance condition in (\ref{p4}), we have
\ben
\hspace{-2cm}\hat I(z)=k\int_{S^1}|\psi_\infty(\hat x,z)|^2d\hat x+k\int_{\Ga_D}\eta(\xi)\left|\psi(\xi,z)+\Im G(\xi,z)\right|^2ds(\xi)+w_{\hat I}(z)\ \ \forall z\in\Om,
\een
where $\|w_{\hat I}\|_{L^\infty(\Om)}\le C(R^{-1}_s+R^{-1}_r)$.
\end{thm}

\debproof By the integral representation we know that
\ben
u^s(x_r,x_s)=\int_{\Ga_D}\left(u^s(\xi,x_s)\frac{\pa G(x_r,\xi)}{\pa\nu(\xi)}-\frac{\pa u^s(\xi,x_s)}{\pa\nu(\xi)}G(x_r,\xi)\right)ds(\xi).
\een
By using Lemma \ref{lem:3.2} we obtain that, for any $z\in\Om$,
\ben
v_b(z,x_s)&=&\int_{\Ga_r}G(z,x_r)\overline{u^s(x_r,x_s)}ds(x_r)\\
&=&\frac 1k\int_{\Ga_D}\Big[u^s(\xi,x_s)\frac{\pa}{\pa\nu}\big(\Im G(\xi,z)+w_r(\xi,z)\big)\\
& &\quad\quad-\frac{\pa u^s(\xi,x_s)}{\pa\nu}\big(\Im G(\xi,z)+w_r(\xi,z)\big)\Big]ds(\xi).
\een
By (\ref{cor3}) we obtain then
\be\label{cor4}
\hat I(z)&=&-k\Im\int_{\Ga_D}\Big[v_s(\xi,z)\frac{\pa}{\pa\nu}\big(\Im G(\xi,z)+w_r(\xi,z)\big)\nn\\
& &\quad\quad-\frac{\pa v_s(\xi,z)}{\pa\nu}\big(\Im G(\xi,z)+w_r(\xi,z)\big)\Big]ds(\xi),
\ee
where $v_s(\xi,z)=k\int_{\Ga_s}G(z,x_s)\overline{u^s(\xi,x_s)}ds(x_s)$. By taking the complex conjugate, we have
\ben
\overline{v_s(\xi,z)}=k\int_{\Ga_s}\overline{G(z,x_s)}u^s(\xi,x_s)ds(x_s).
\een
Thus $\overline{v_s(\xi,z)}$ is a weighted superposition of the scattered waves $u^s(\xi,x_s)$. Therefore, $\overline{v_s(\xi,z)}$ is the radiation solution of the Helmholtz equation
\ben
\De_\xi\overline{v_s(\xi,z)}+k^2\overline{v_s(\xi,z)}=0\ \ \ \ \mbox{in }\R^2\bks\bar D
\een
satisfying the impedance boundary condition
\ben
\fl \left(\frac{\pa}{\pa\nu(\xi)}+\i k\eta(\xi)\right)\overline{v_s(\xi,z)}
&=&k\int_{\Ga_s}\overline{G(z,x_s)}\left(\frac{\pa}{\pa\nu(\xi)}+\i k\eta(\xi)\right)u^s(\xi,x_s)ds(x_s)\\
&=&-k\int_{\Ga_s}\overline{G(z,x_s)}\left(\frac{\pa}{\pa\nu(\xi)}+\i k\eta(\xi)\right)G(\xi,x_s)ds(x_s)\\
&=&-\left(\frac{\pa}{\pa\nu(\xi)}+\i k\eta(\xi)\right)\big(\Im G(\xi,z)+w_s(\xi,z)\big)\ \ \ \ \mbox{on }\Ga_D,
\een
where we have used Lemma \ref{lem:3.2} in the last equality. This implies by using (\ref{ps1})-(\ref{ps2}) that $\overline{v_s(\xi,z)}=\psi(\xi,z)+\zeta(\xi,z)$, where
$\zeta$ satisfies the impedance scattering problem in Lemma \ref{lem:3.5} with $g(\cdot)=\left(\frac{\pa }{\pa\nu}+\i k\eta\right)w_s(\cdot,z)$.
By the Lemma \ref{lem:3.2} and Lemma {\ref{lem:3.5}}, $\zeta$ satisfies $\|\zeta(\cdot,z)\|_{H^{1/2}(\Ga_D)}\le CR^{-1}_s$ uniformly for $z\in\Om$. Now from the boundary condition satisfied
by $\zeta$ on $\Ga_D$ we know also that $\|\frac{\pa\zeta(\cdot,z)}{\pa\nu}\|_{H^{-1/2}(\Ga_D)}\le CR_s^{-1}$ uniformly for $z\in\Om$.

Now substituting $v_s(\xi,z)=\overline{\psi(\xi,z)}+\overline{\zeta(\xi,z)}$ into (\ref{cor4}) we obtain
\ben
\fl \hat I(z)
&=&-\,\Im\int_{\Ga_D}\left(\overline{\psi(\xi,z)}\frac{\pa\Im G(\xi,z)}{\pa\nu(\xi)}-\frac{\pa\overline{\psi(\xi,z)}}{\pa\nu(\xi)}\Im G(\xi,z)\right)ds(\xi)+O(R^{-1}_s+R^{-1}_r)\\
\fl&=&\,\Im\int_{\Ga_D}\left(\psi(\xi,z)\frac{\pa\Im G(\xi,z)}{\pa\nu(\xi)}-\frac{\pa\psi(\xi,z)}{\pa\nu(\xi)}\Im G(\xi,z)\right)ds(\xi)+O(R^{-1}_s+R^{-1}_r).
\een
By (\ref{ps2}) we have
\ben
 \fl \ \ \ \ \Im\int_{\Ga_D}\left(\psi(\xi,z)\frac{\pa\Im G(\xi,z)}{\pa\nu(\xi)}-\frac{\pa\psi(\xi,z)}{\pa\nu(\xi)}\Im G(\xi,z)\right)ds(\xi)\\
\fl =\Im\int_{\Ga_D}\Big[\psi(\xi,z)\left(\frac{\pa\Im G(\xi,z)}{\pa\nu(\xi)}-\i k\eta(\xi)\Im G(\xi,z)\right)\\
\fl \qquad-\left(\frac{\pa\psi(\xi,z)}{\pa\nu(\xi)}+\i k\eta(\xi)\psi(\xi,z)\right)\Im G(\xi,z)+2\i k\eta(\xi)\Im G(\xi,z)\psi(\xi,z)\Big]ds(\xi)\\
\fl=\Im\int_{\Ga_D}\Big[-\psi(\xi,z)\cdot \left(\frac{\pa\overline{\psi(\xi,z)}}{\pa\nu(\xi)}-\i k\eta(\xi)\overline{\psi(\xi,z)}\right)\\
\fl \qquad+\left(\frac{\pa\Im G(\xi,z)}{\pa\nu(\xi)}+\i k\eta(\xi)\Im G(\xi,z)\right)\Im G(\xi,z)+2\i k\eta(\xi)\Im G(\xi,z)\psi(\xi,z)\Big]ds(\xi)\\
\fl=-\,\Im\la\psi(\cdot,z),\Lam_D(\psi(\cdot,z))\ra_{\Ga_D}+k\int_{\Ga_D}\eta(\xi)\left|\psi(\xi,z)+\Im G(\xi,z)\right|^2ds(\xi).
\een
This completes the proof by using Lemma \ref{lem:3.3}.
\finproof

For the ease of later reference, we state the results for the sound soft non-penetrable obstacles in the following theorem.

\begin{thm}\label{thm:3.3}
For any $z\in\Om$, let $\psi(x,z)$ be the radiation solution of the Helmholtz equation
\ben
& &\De\psi(x,z)+k^2\psi(x,z)=0\ \ \ \ \mbox{in }\R^2\bks\bar D, \\
& &\psi(x,z)=-\Im G(x,z)\ \ \ \ \mbox{on }\Ga_D.
\een
Then if the measured field $u^s=u-u^i$ with $u$ satisfying the problem (\ref{p3})-(\ref{p5}) with the Dirichlet condition in (\ref{p4}), we have
\ben
\hat I(z)=k\int_{S^1}|\psi_\infty(\hat x,z)|^2d\hat x+w_{\hat I}(z)\ \ \ \ \forall z\in\Om,
\een
where $\|w_{\hat I}\|_{L^\infty(\Om)}\le C(R^{-1}_s+R^{-1}_r)$.
\end{thm}

We remark that for the non-penetrable obstacles, $\psi(x,z)$ is again the scattering solution of the Helmholtz equation with the incident wave $\Im G(x,z)$. Similar to the remark after the proof of Theorem \ref{thm:3.1}, we expect that the imaging functional will have contrast at the boundary of the scatterer $D$ and decay away from the scatterer.

\section{Numerical experiments}\label{section4}

In this section, we show a variety of numerical experiments to illustrate the imaging power of the RTM algorithm proposed
in this paper. To synthesize the scattering data we compute the solution $u(x,x_s)$ of the scattering problem (\ref{p1})-(\ref{p2}) or (\ref{p3})-(\ref{p5}) by standard Nystr\"{o}m's methods \cite{colton-kress}. The boundary integral equations on $\Ga_D$ are solved on a uniform mesh over the boundary with ten points per probe wavelength. The boundaries of the obstacles used in our numerical experiments are parameterized as follows:
\ben
\mbox{Circle:}\ \ \ \ &&x_1=\rho\cos(\theta),\ \ x_2=\rho\sin(\theta),\ \ \theta\in (0,2\pi],\\
\mbox{Kite:}\ \ \ \ &&x_1=\cos(\theta) + 0.65\cos(2\theta) - 0.65,\ \ x_2=1.5 \sin (\theta),\ \ \theta\in (0,2\pi],\\
\mbox{$p$-leaf:}\ \ \ \ &&r(\theta)=1+0.2\cos(p\theta),\ \ \theta\in (0,2\pi].
\een

\bigskip
\textbf{Example 1}.
In this example we consider the imaging of a circle of radius $\rho=2$ centered at the origin. We compare the results of the imaging functional in (\ref{cor2}) with the corresponding theoretical results in Theorems \ref{thm:3.1} and \ref{thm:3.3}. Let $\Ga_r=\Ga_s$ be the circle of radius $R=10$.
Let $\Om=(-3,3)\times(-3,3)$ be the search region and the imaging functional is computed at the nodal points of a uniform $201\times 201$ mesh.
We test two probe wavelengths $\lambda= 1$ and $\lam=0.25$, where $\lam=2\pi/k$.

\begin{figure}
    \centering
    \includegraphics[width=1\textwidth]{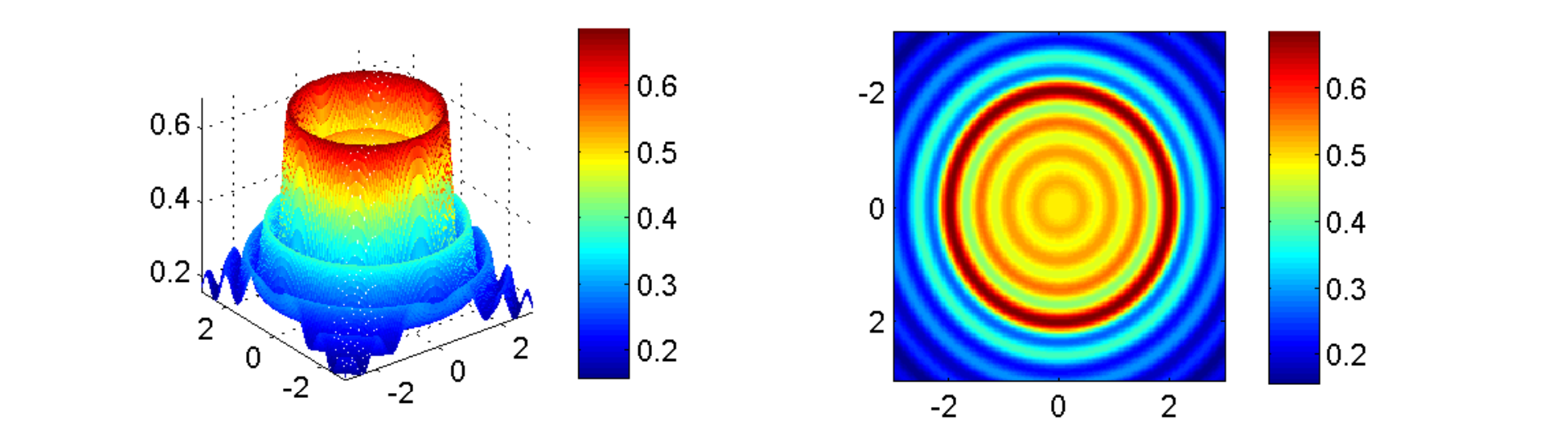}
    \includegraphics[width=1\textwidth]{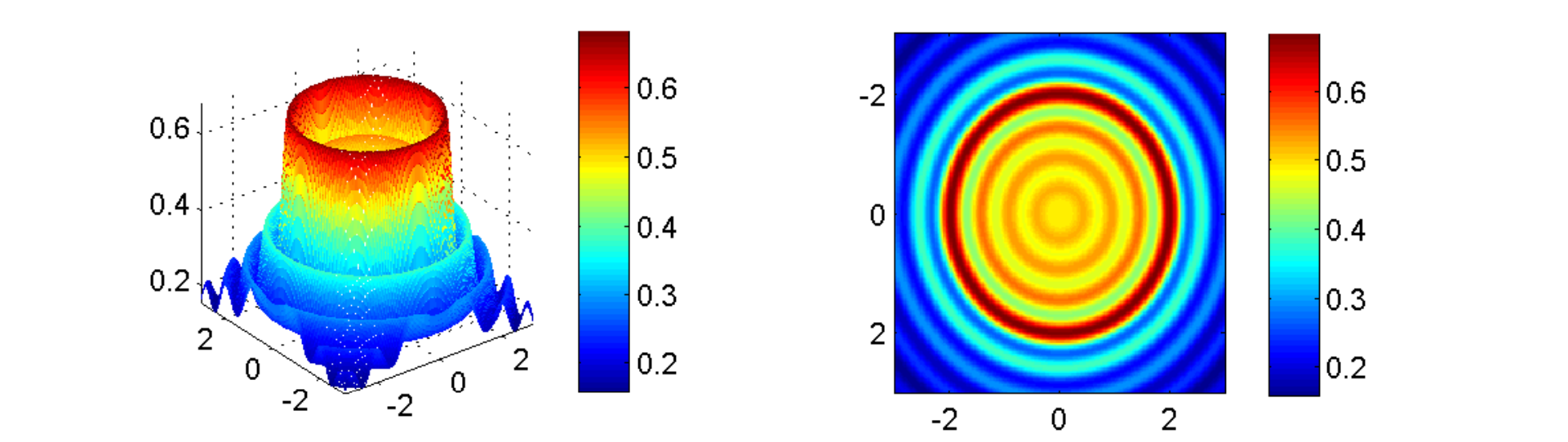}
\caption{Example 1: Non-penetrable obstacle with Dirichlet condition, probe wavelength $\lambda=1$, and $N_s=N_r=64$.
The first row shows the surface plot and the contour plot of the imaging functional in Theorem \ref{thm:3.3} (ignoring $w_{\hat I}$). The second row shows the surface plot and the contour plot of the imaging functional in (\ref{cor2}).}  \label{figure_1}
\end{figure}

\begin{figure}
    \centering
    \includegraphics[width=1\textwidth]{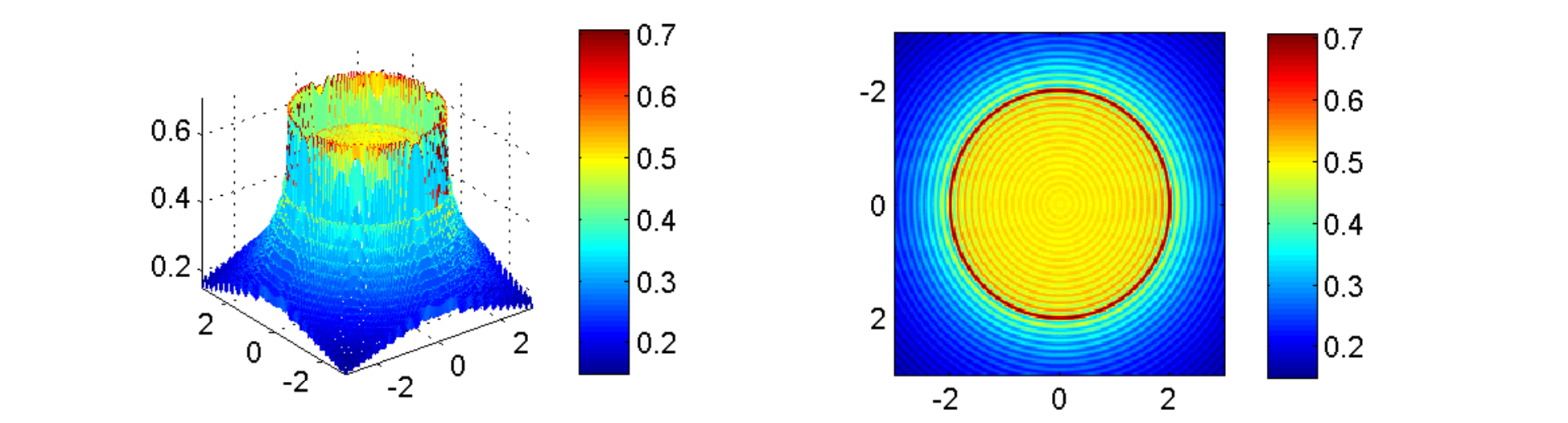}
    \includegraphics[width=1\textwidth]{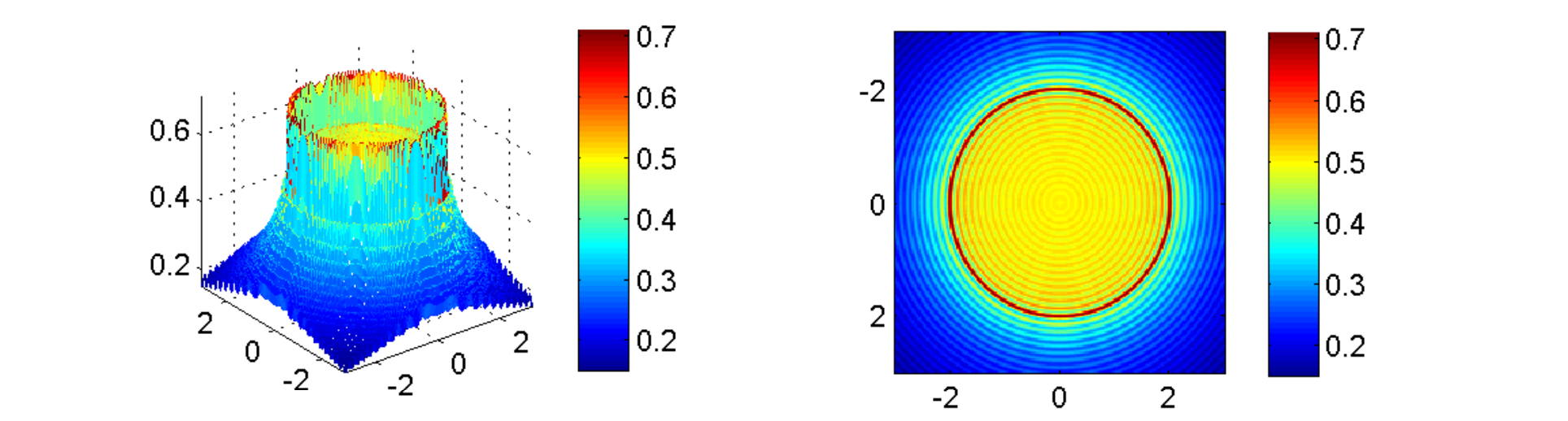}
\caption{Example 1: Non-penetrable obstacle with Dirichlet condition, probe wavelength $\lambda=0.25$, and $N_s=N_r=256$. The first row shows the surface plot and the contour plot of the imaging functional in Theorem \ref{thm:3.3} (ignoring $w_{\hat I}$). The second row shows the surface plot and the contour plot of the imaging functional in (\ref{cor2}).} \label{figure_2}
\end{figure}

\begin{figure}
    \centering
    \includegraphics[width=1\textwidth]{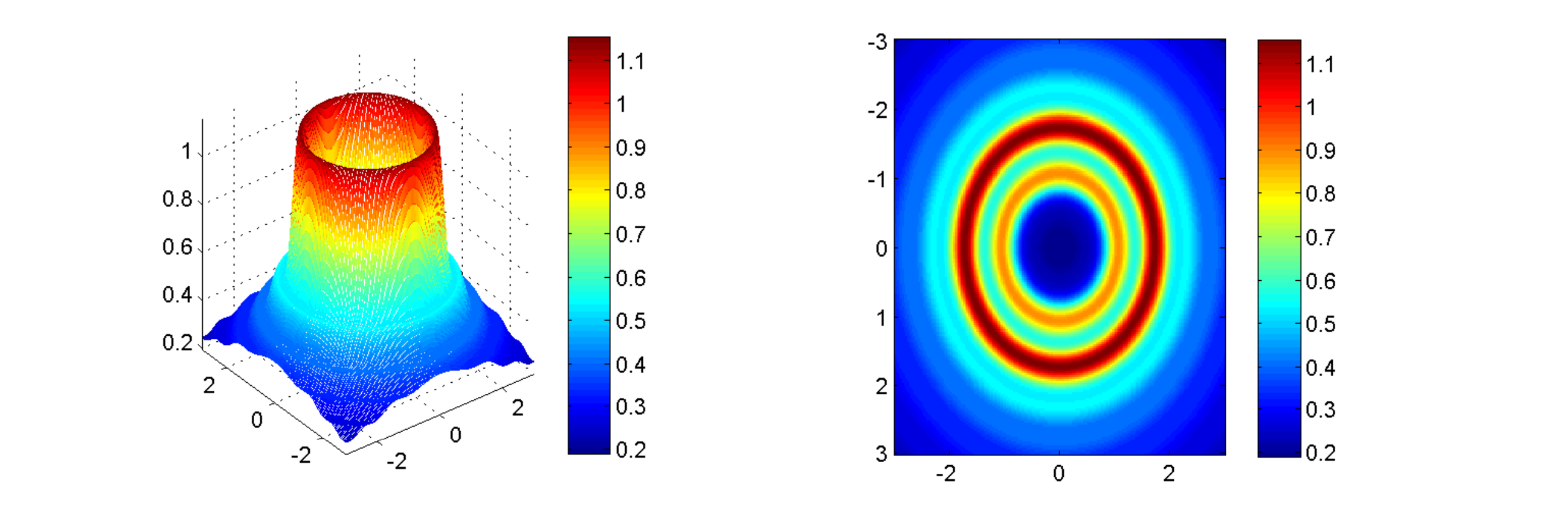}
   \includegraphics[width=1\textwidth]{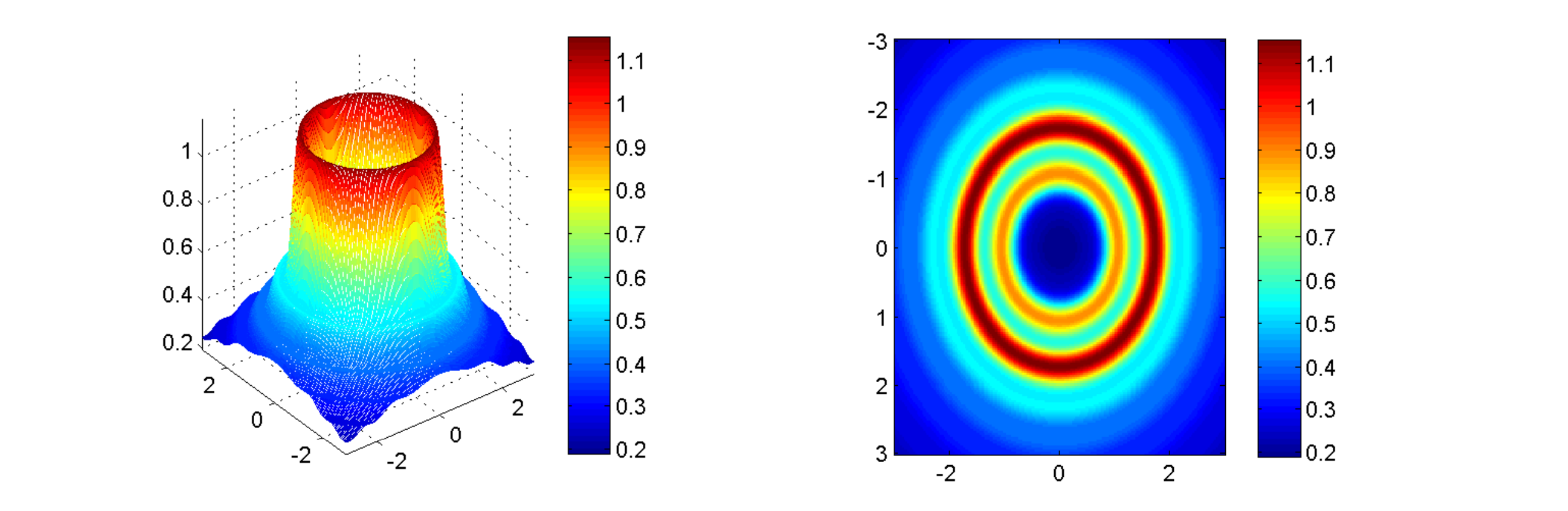}
\caption{Example 1: Penetrable obstacles, probe wavelength $\lambda=1$, and $N_s=N_r=64$. The first row shows the surface plot and the contour plot of the imaging functional in Theorem \ref{thm:3.3} (ignoring $w_{\hat I}$). The second row shows the surface
plot and the contour plot of the imaging functional in (\ref{cor2}).} \label{figure_3}
\end{figure}

\begin{figure}
    \centering
    \includegraphics[width=1\textwidth]{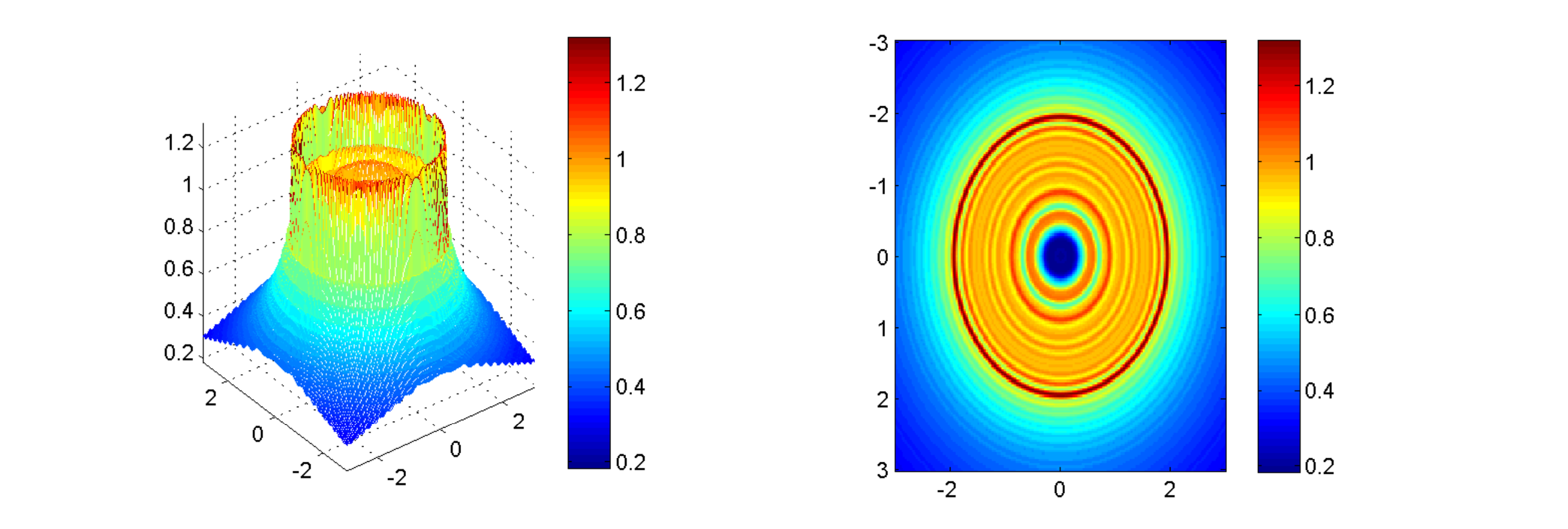}
    \includegraphics[width=1\textwidth]{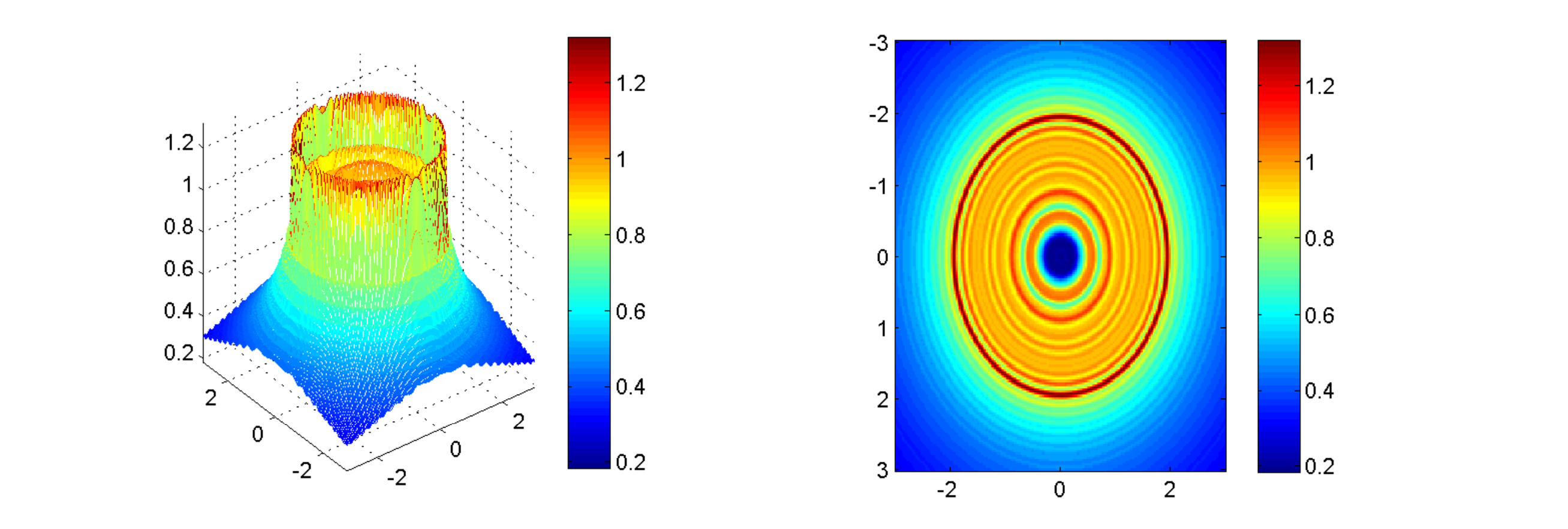}
\caption{Example 1: Penetrable obstacles, probe wavelength $\lambda=0.25$, and $N_s=N_r=256$. The first row shows the surface plot and the contour plot of the imaging functional in Theorem \ref{thm:3.3} (ignoring $w_{\hat I}$). The second row shows the surface
plot and the contour plot of the imaging functional in (\ref{cor2}).} \label{figure_4}
\end{figure}

Figure \ref{figure_1} and Figure \ref{figure_2} show the comparison of the imaging functional for a non-penetrable obstacle with Dirichlet condition on $\Ga_D$. Figure \ref{figure_3} and Figure \ref{figure_4} show the results for a penetrable obstacle with $n(x)=0.25$. We observe that the imaging functional (\ref{cor2}) agrees well with the theoretical results in Theorems \ref{thm:3.1} and \ref{thm:3.3}.

\begin{figure}
    \centering
    \includegraphics[width=0.4\textwidth]{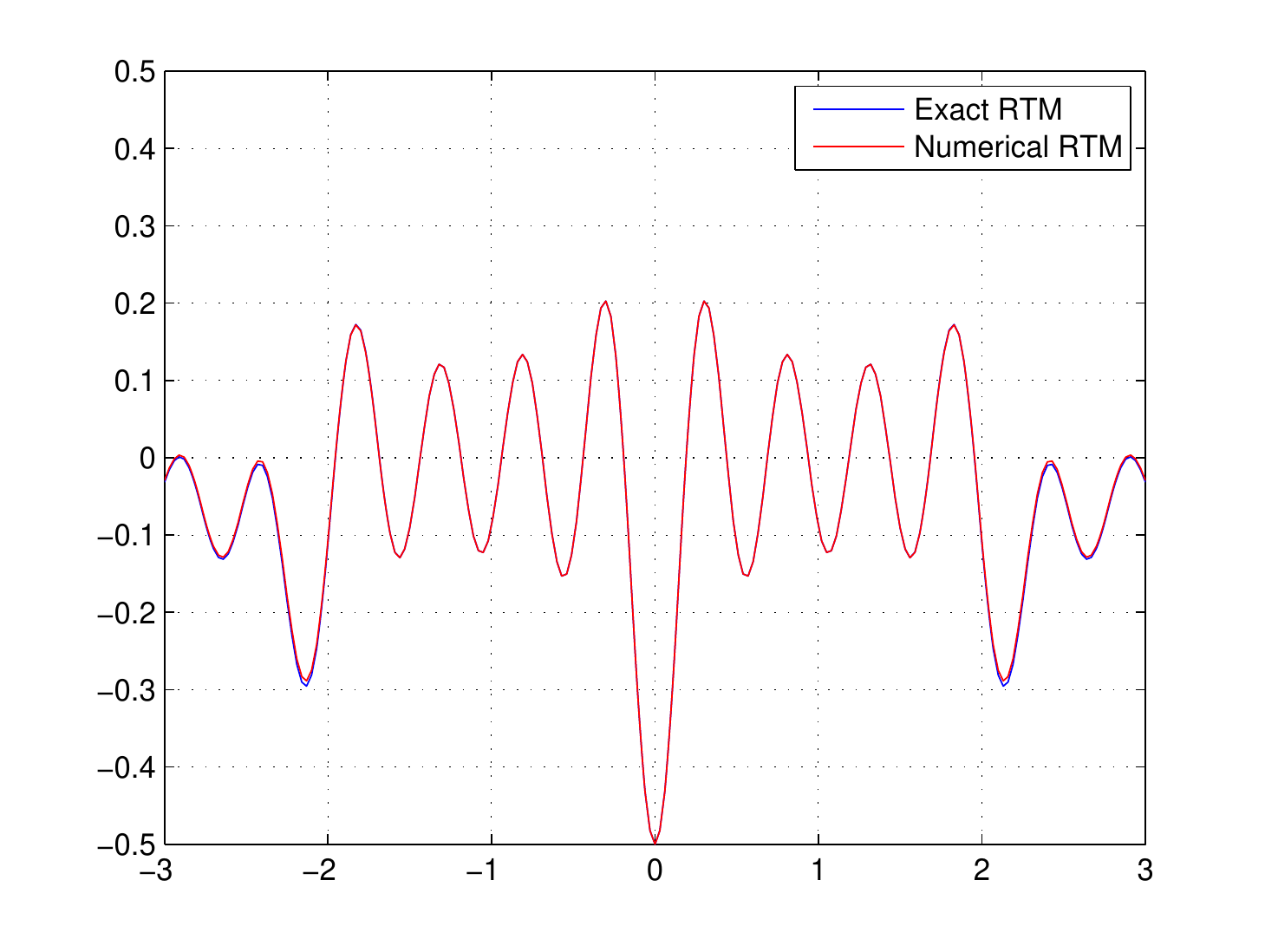}
    \includegraphics[width=0.4\textwidth]{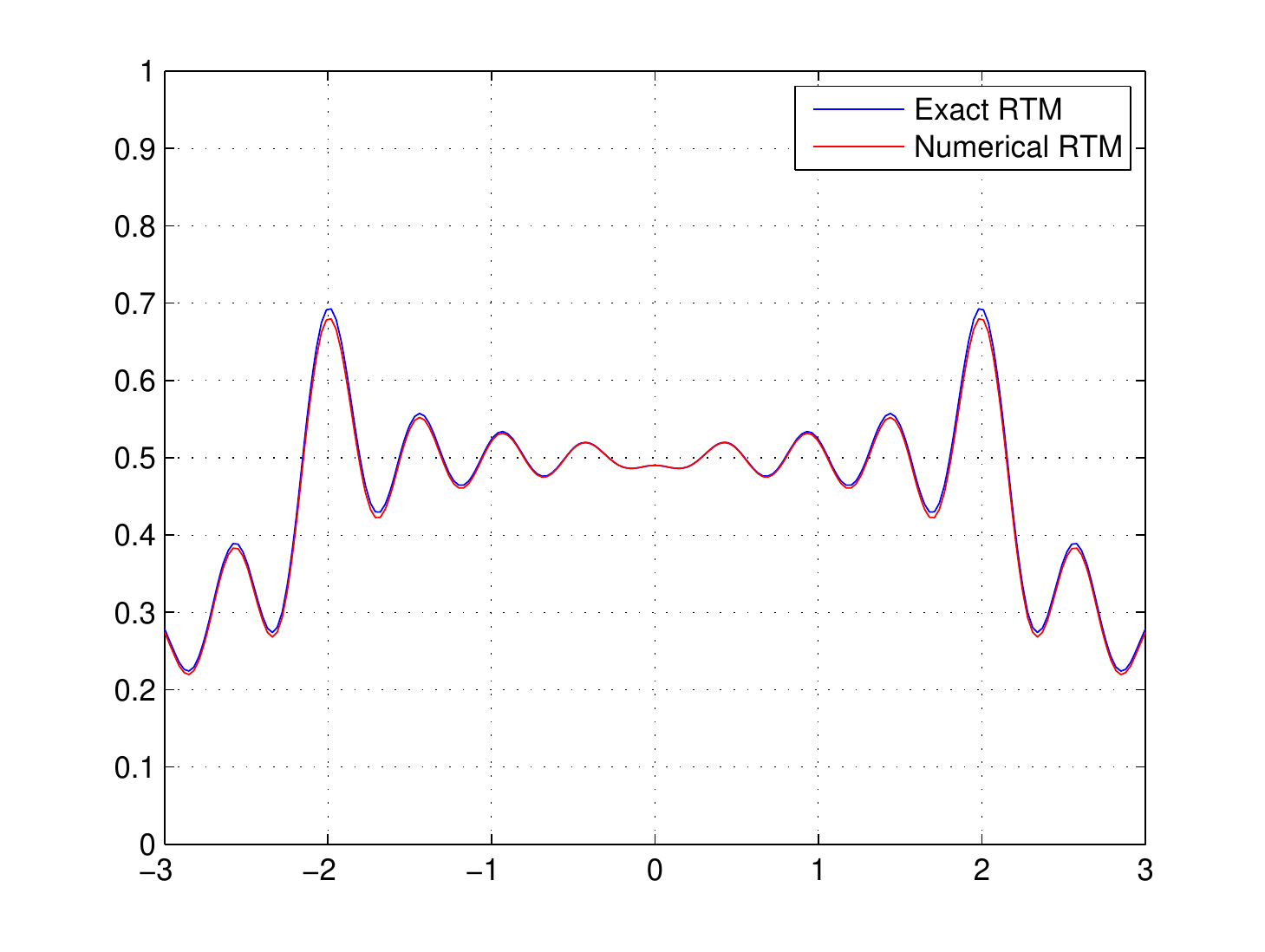}
     \includegraphics[width=0.4\textwidth]{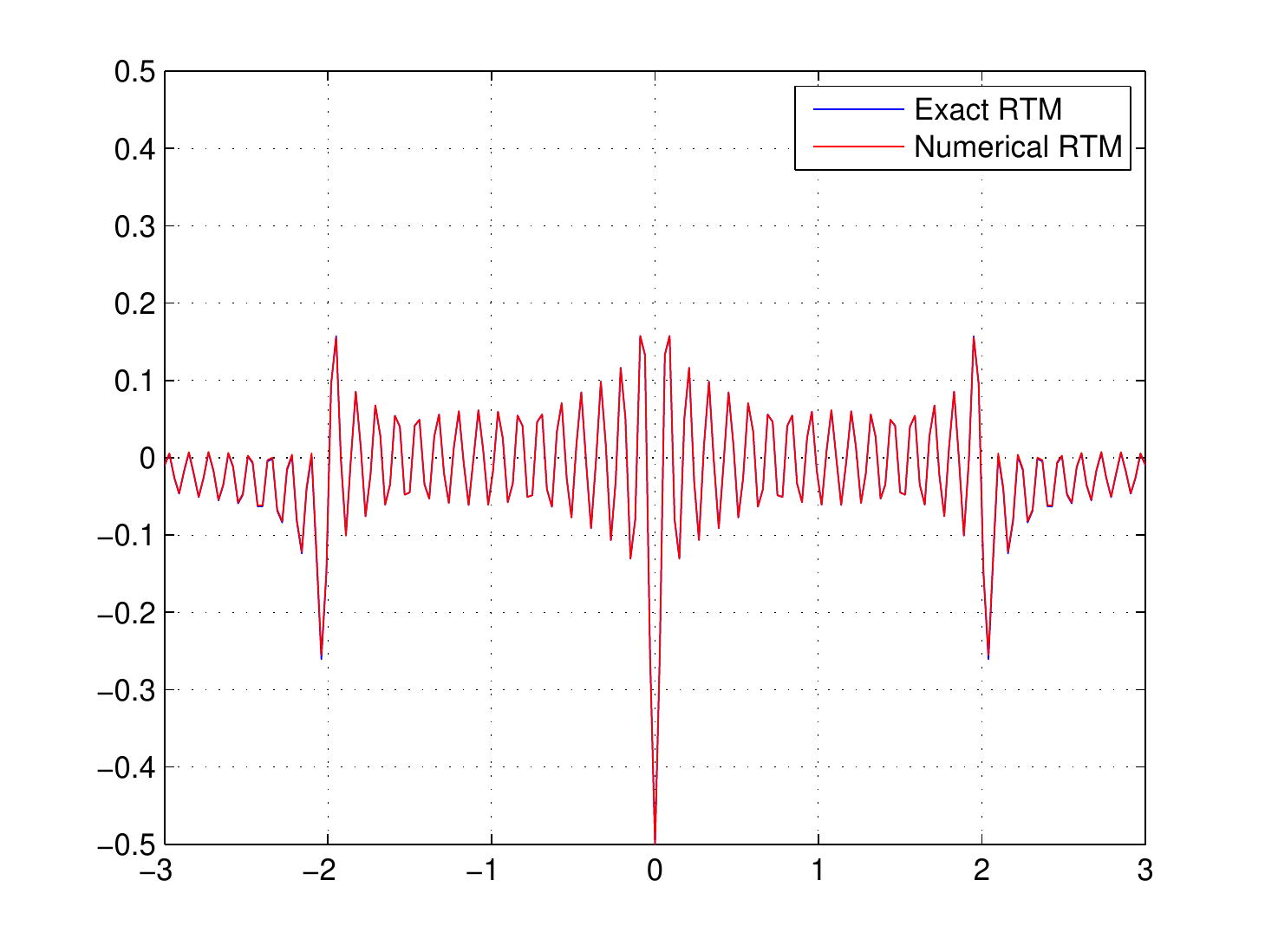}
    \includegraphics[width=0.4\textwidth]{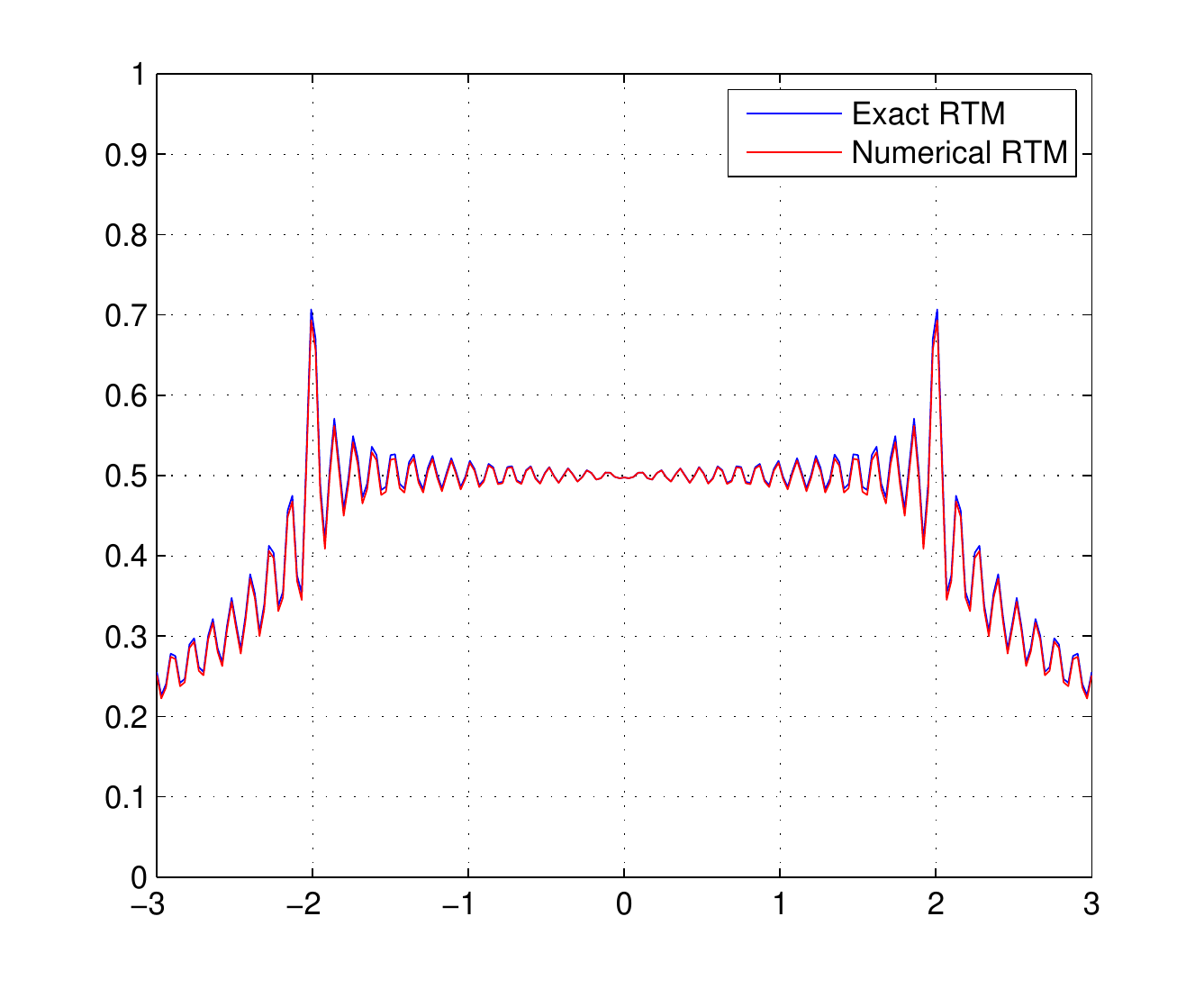}
    \includegraphics[width=0.4\textwidth]{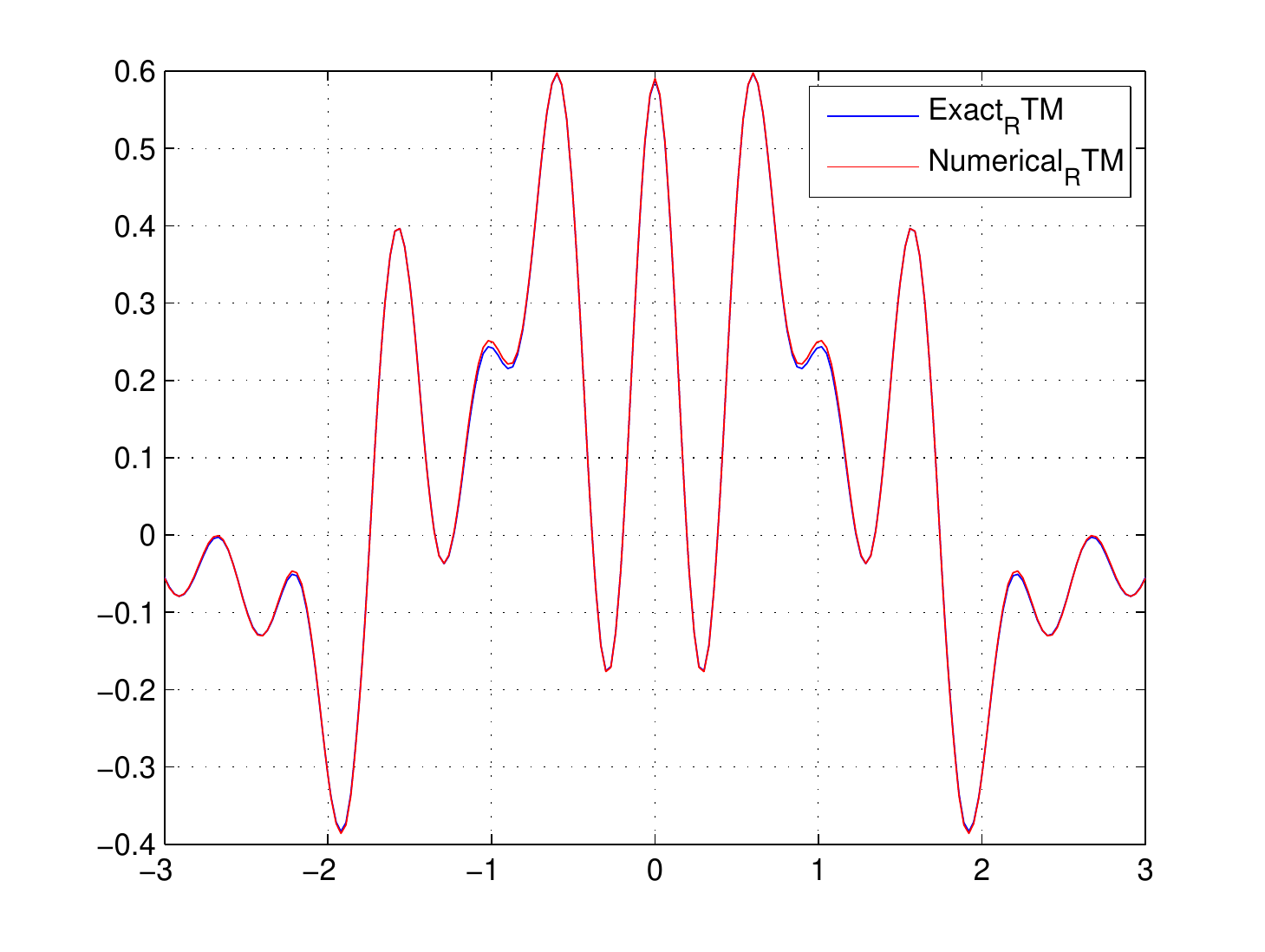}
    \includegraphics[width=0.4\textwidth]{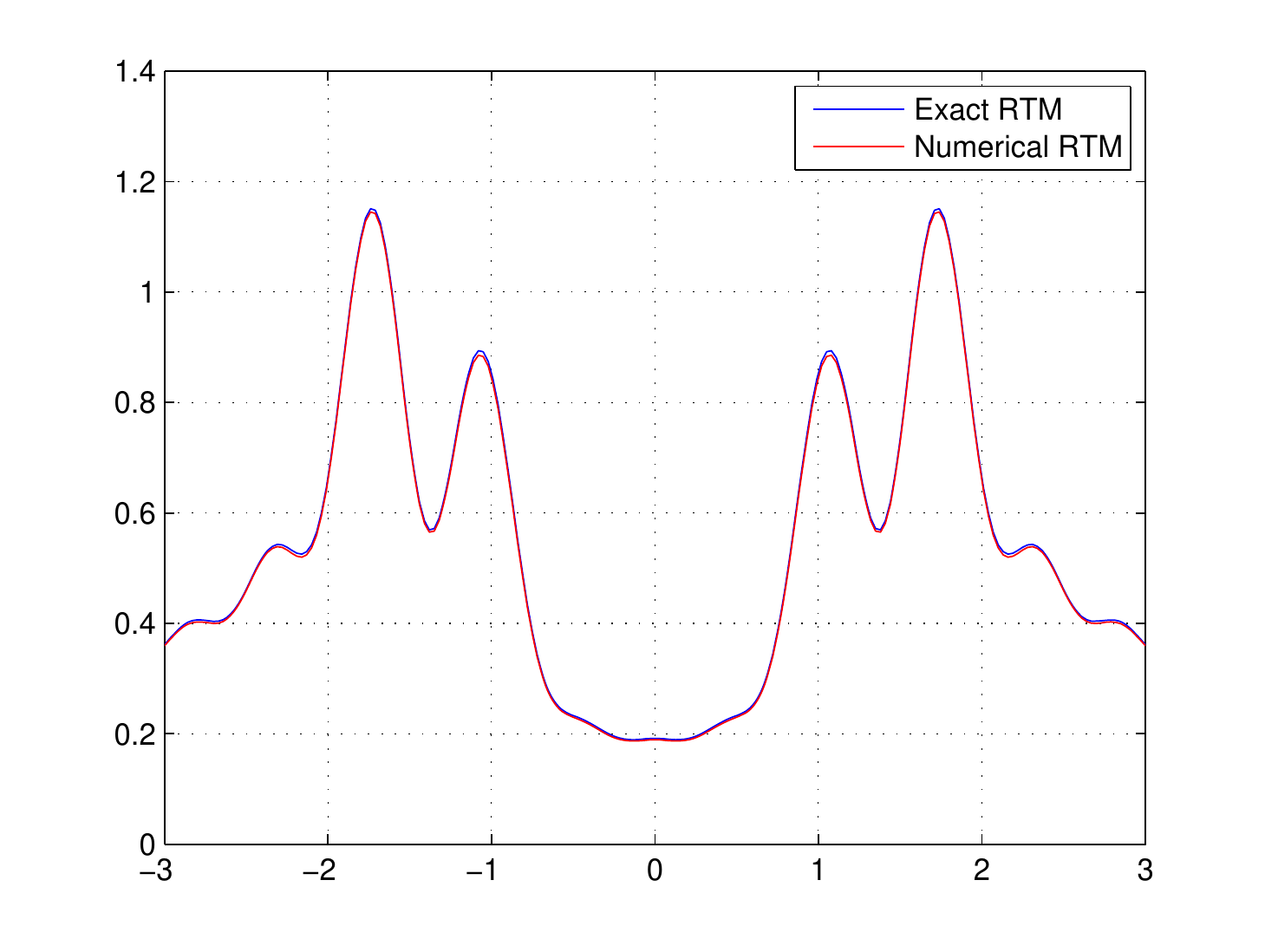}
    \includegraphics[width=0.4\textwidth]{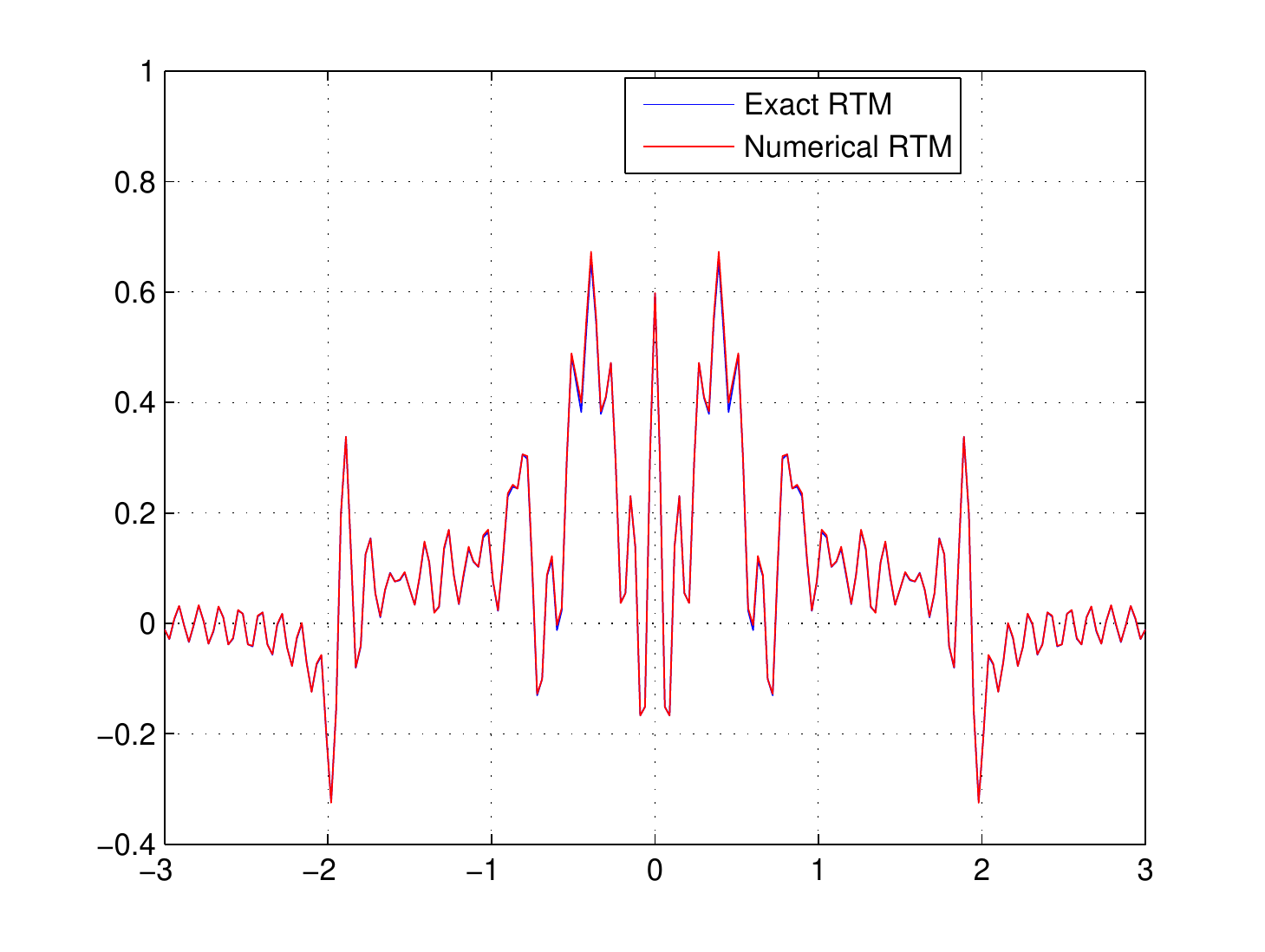}
    \includegraphics[width=0.4\textwidth]{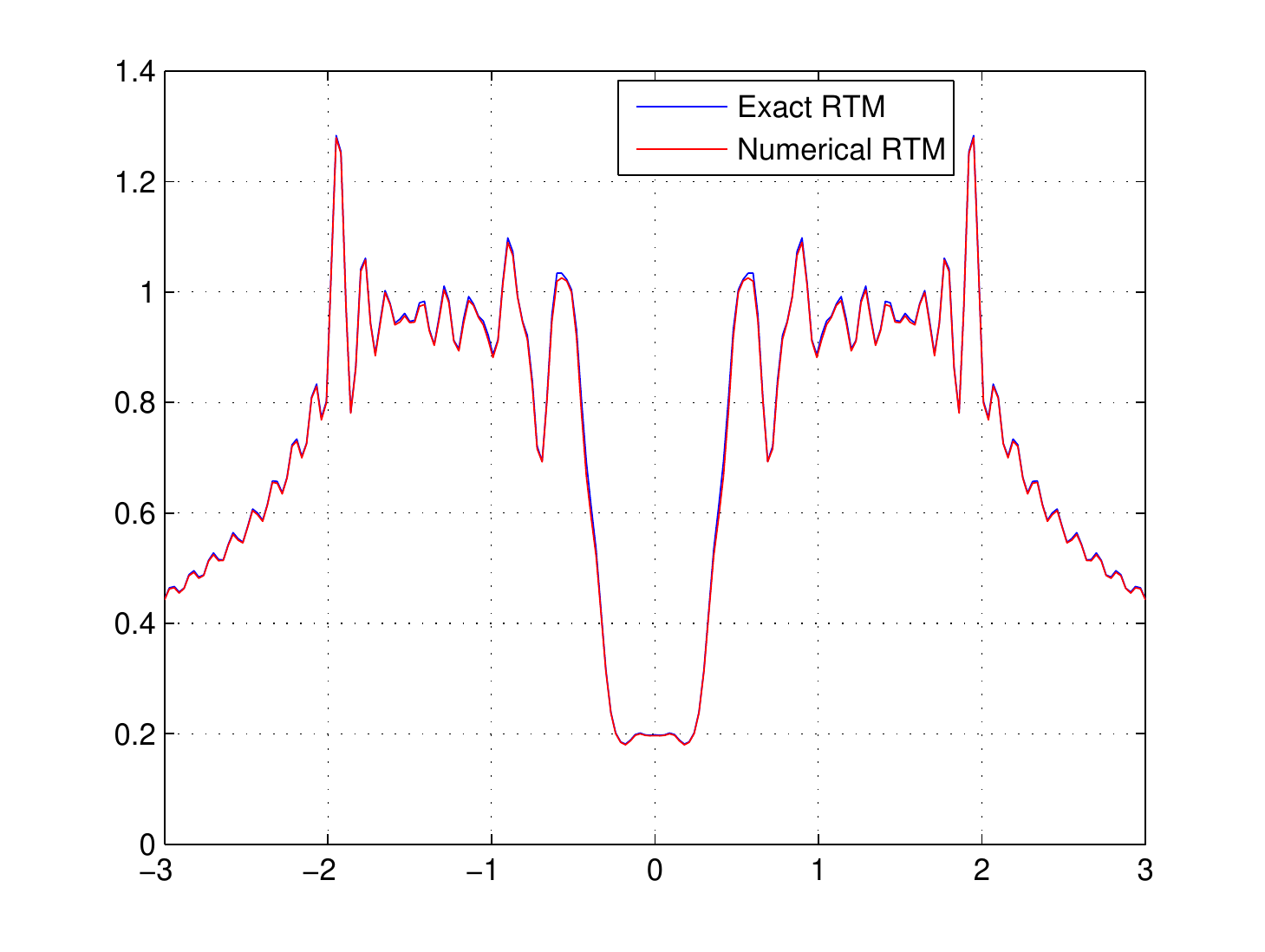}
    \caption{Example 1: The comparison of the real (left) and imaginary (right) part of the cross-correlation functional. The first and second row is for the non-penetrable obstacle with Dirichlet condition and the third and fourth row is for the penetrable obstacle with $n(x)=0.25$. The first and third row show the cross sections at $x_1=0$ of the real and imaginary part of the cross-correlation functional when $\lam=1$ and $N_s=N_r=64$. The second and fourth row show the cross sections at $x_1=0$ of the real and imaginary part of the cross-correlation functional when $\lambda=0.25$ and $N_s=N_r=256$. } \label{figure_real}
\end{figure}

Figure \ref{figure_real} shows the comparison of the real part of the cross-correlation functional
\ben
\fl \qquad \tilde I(z)=-k^2\cdot\Re\left\{\frac{|\Ga_s||\Ga_r|}{N_sN_r}\sum^{N_s}_{s=1}\sum^{N_r}_{r=1}G(z,x_s)G(z,x_r)\overline{u^s(x_r,x_s)}\right\}\ \ \ \ \forall z\in\Om.
\een
and the imaginary part of the correlation functional in (\ref{cor2}). We observe that the imaging quality, i.e., the contrast near the boundary of the scatterer, of $\tilde I(z)$ is worse than that of the imaginary part of the cross-correlation functional (\ref{cor2}).

\begin{figure}
    \centering
    \includegraphics[width=1\textwidth]{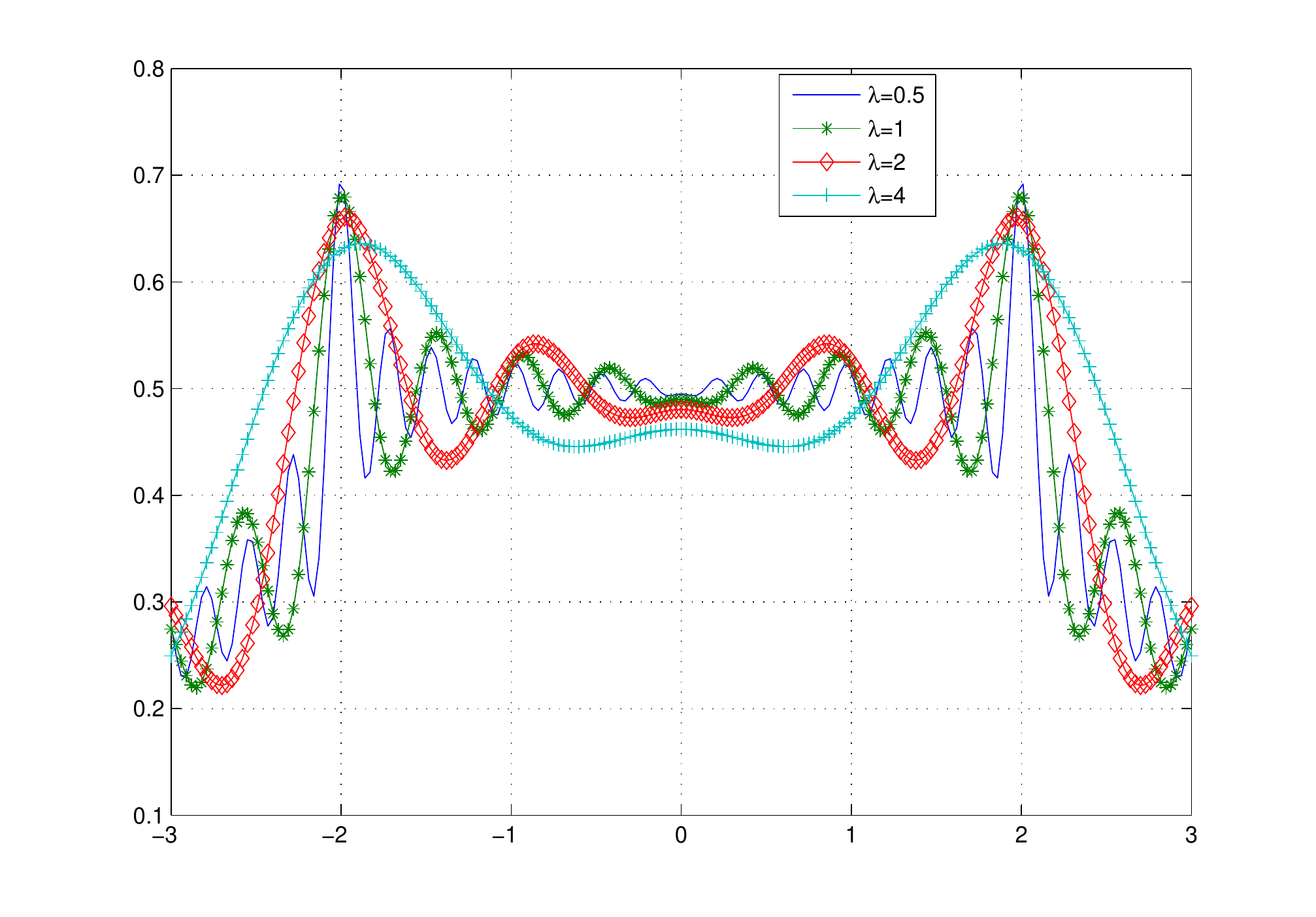}
\caption{Example 1: Comparison of cross-sections of the imaging functional at $x_1=0$ for different probe wavelengths $\lambda=4,2,1,0.5$.} \label{figure_comparison}
\end{figure}

Figure \ref{figure_comparison} depicts cross-sections of the imaging functional at $x_1=0$ for probe wavelengths $\lambda=4,2,1,0.5$.
It shows clearly the resolution is improved with the increase of the wavenumber.


\begin{figure}
 \centering
    \includegraphics[width=0.24\textwidth, height=1.5in]{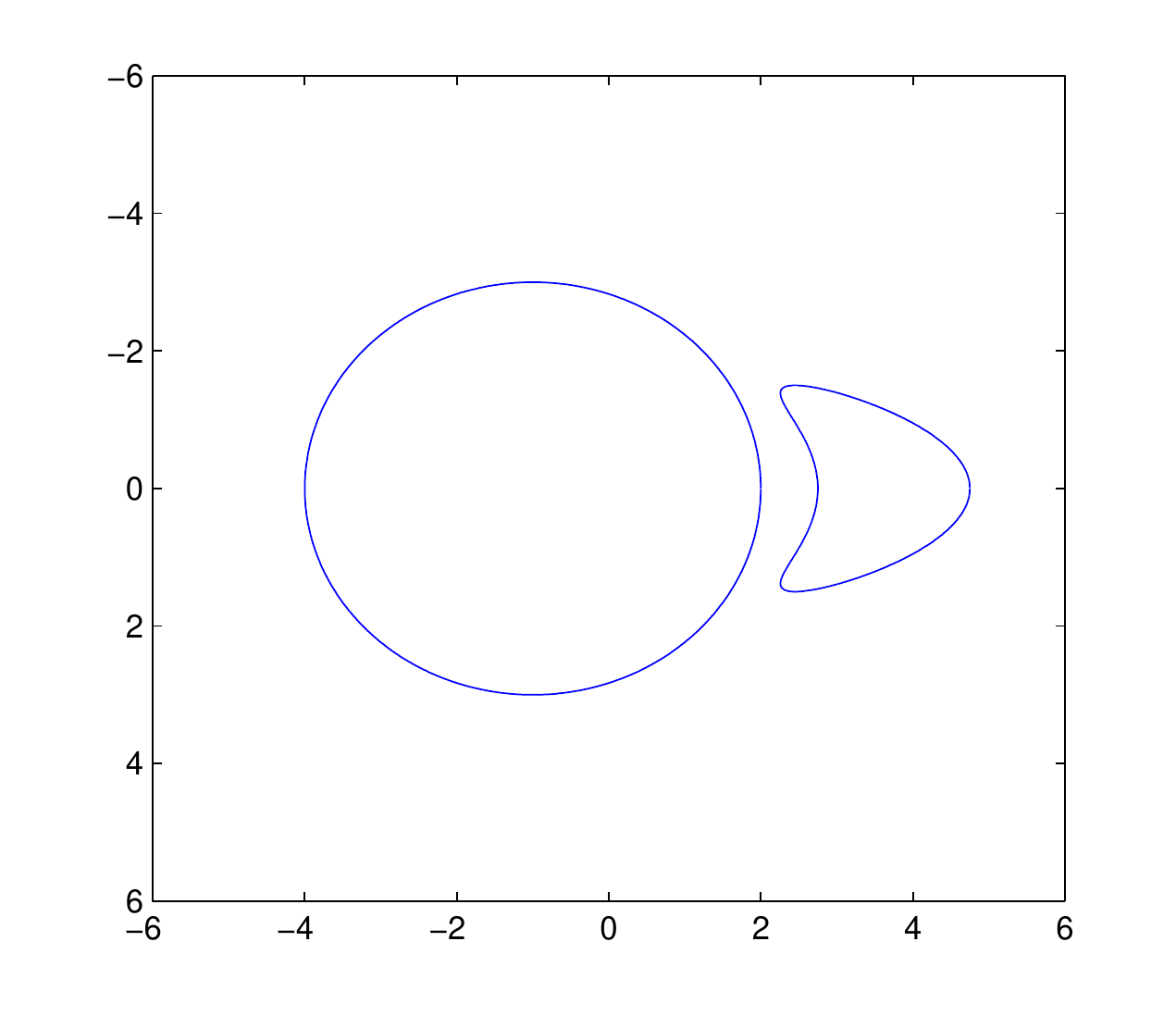}
    \includegraphics[width=0.24\textwidth, height=1.5in]{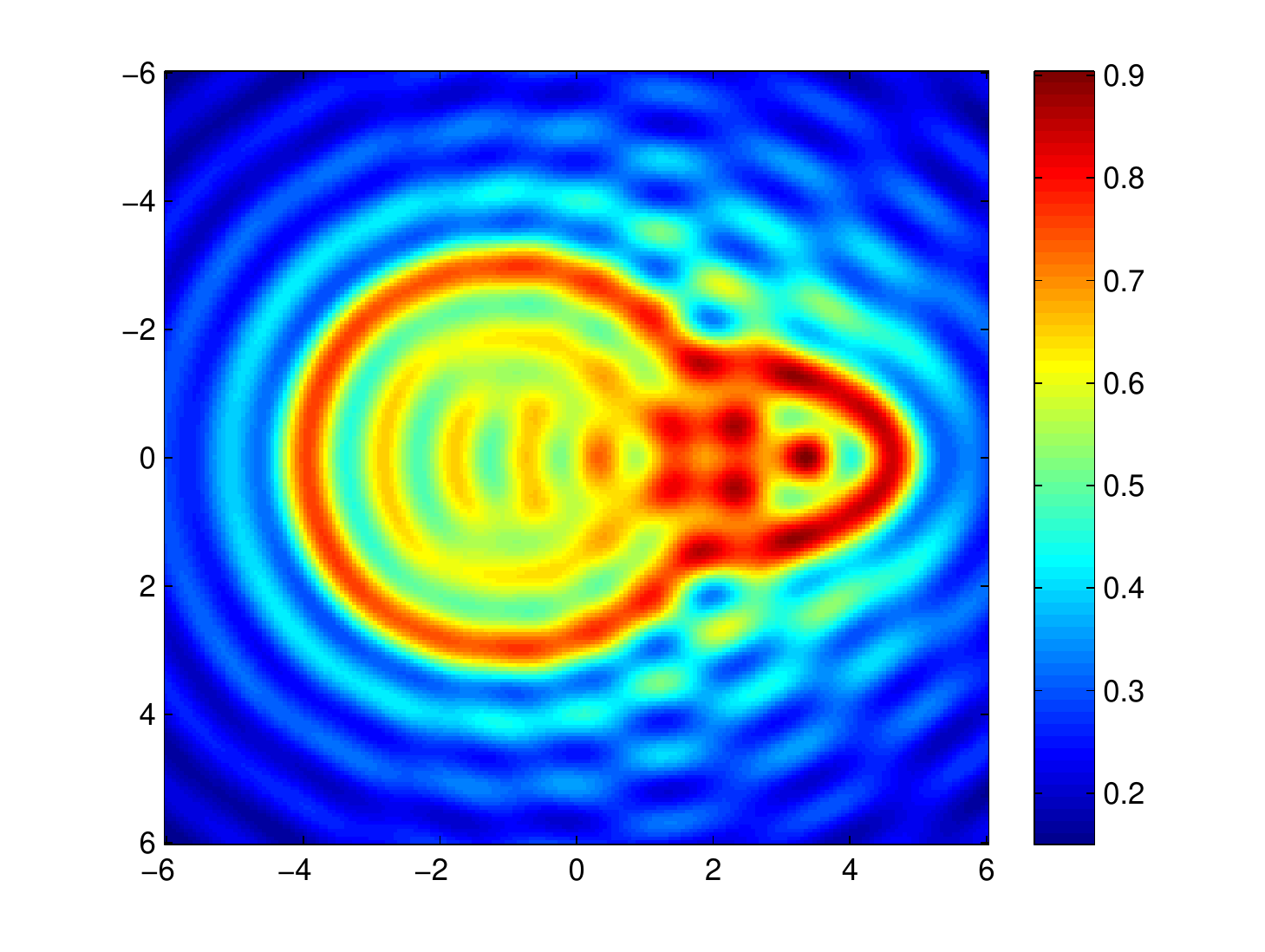}
    \includegraphics[width=0.24\textwidth, height=1.5in]{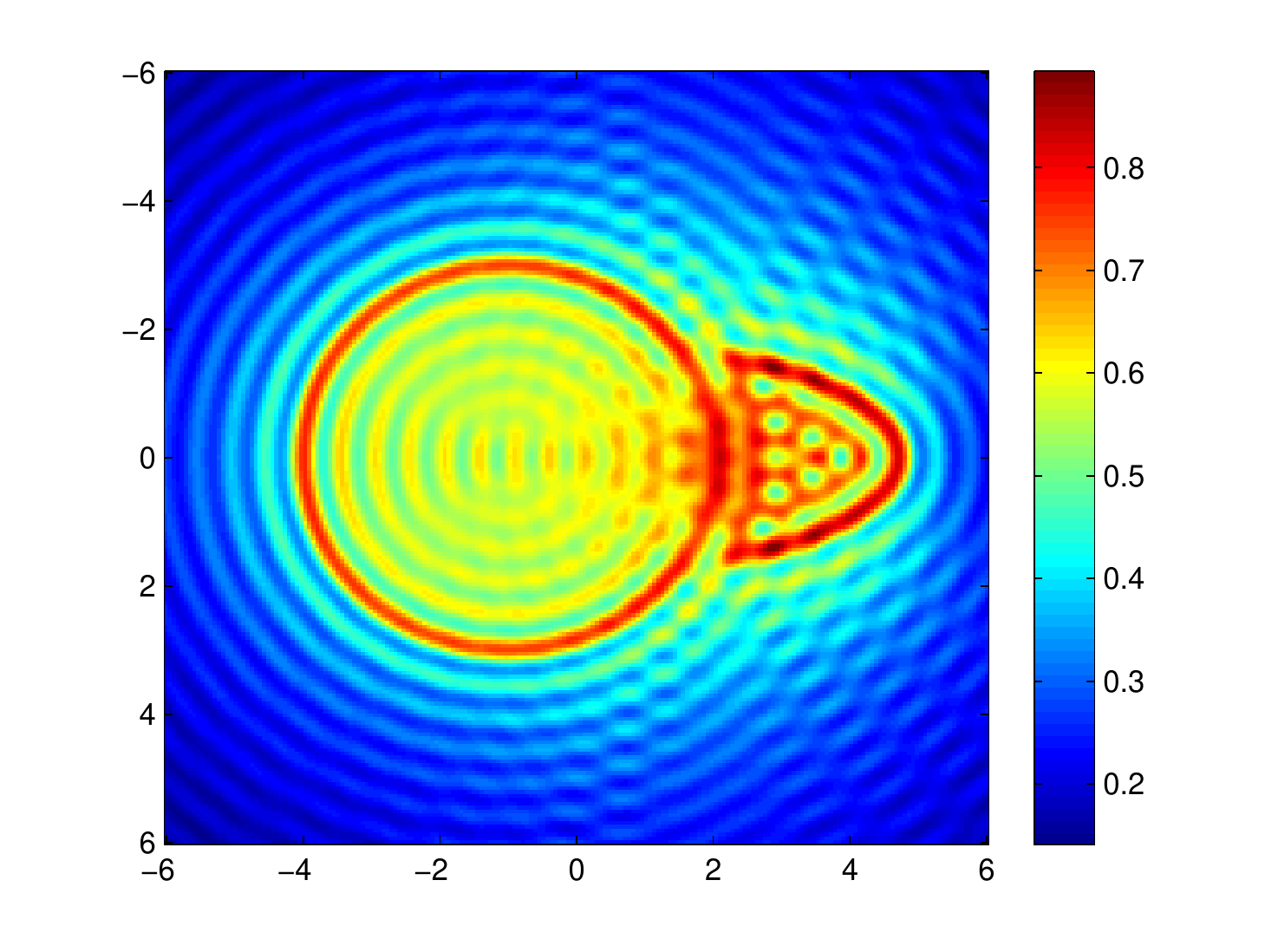}
    \includegraphics[width=0.24\textwidth, height=1.5in]{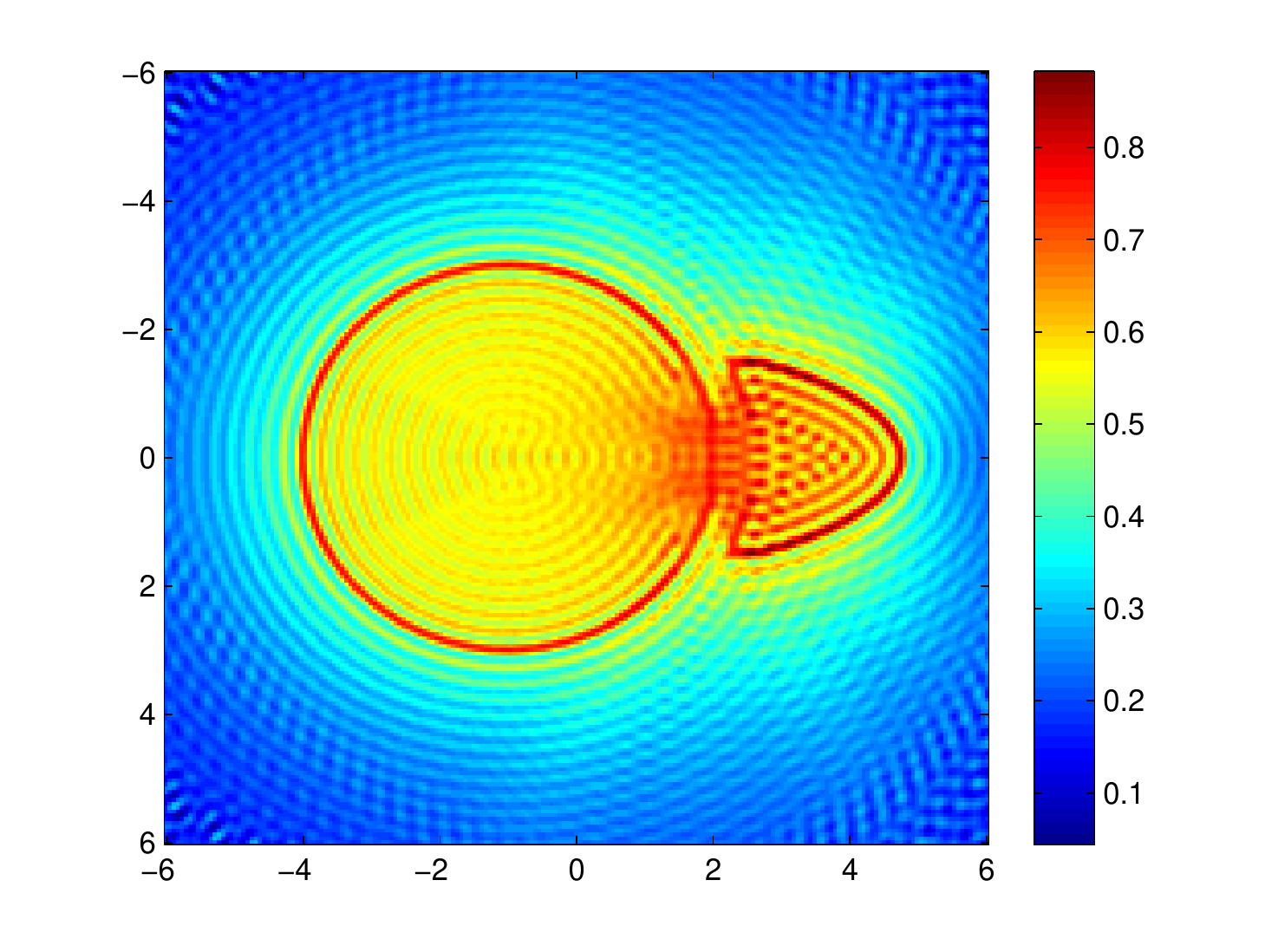}
    \caption{Example 2: The first picture is the exact obstacles. The other three pictures from left to right are imaging results using probe wavelengths $\lambda=2,1,0.5$ and $N_s=N_r=128$, respectively. } \label{figure_6}
\end{figure}

\begin{figure}
 \centering
    \includegraphics[width=0.3\textwidth, height=1.5in]{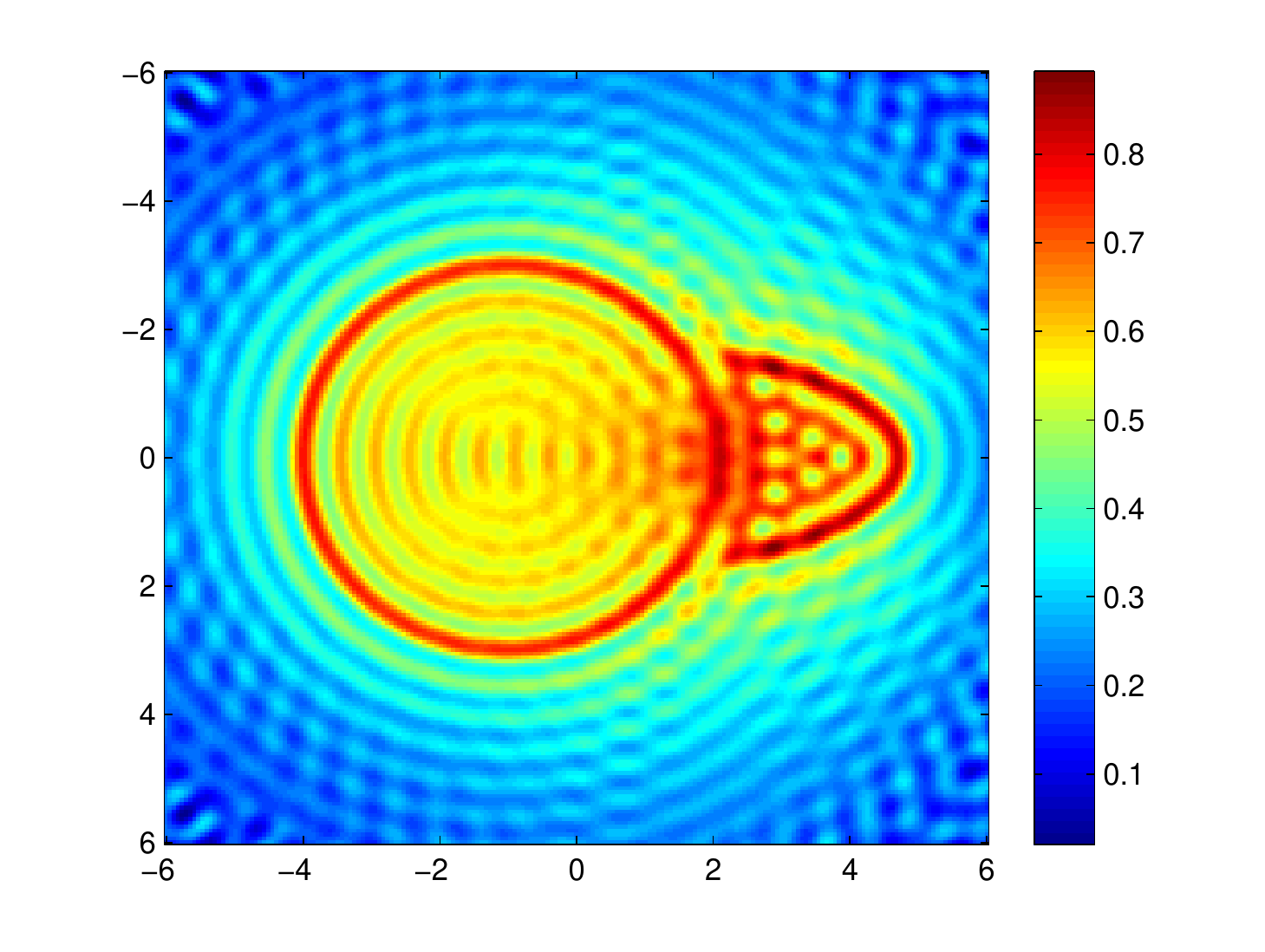}
    \includegraphics[width=0.3\textwidth, height=1.5in]{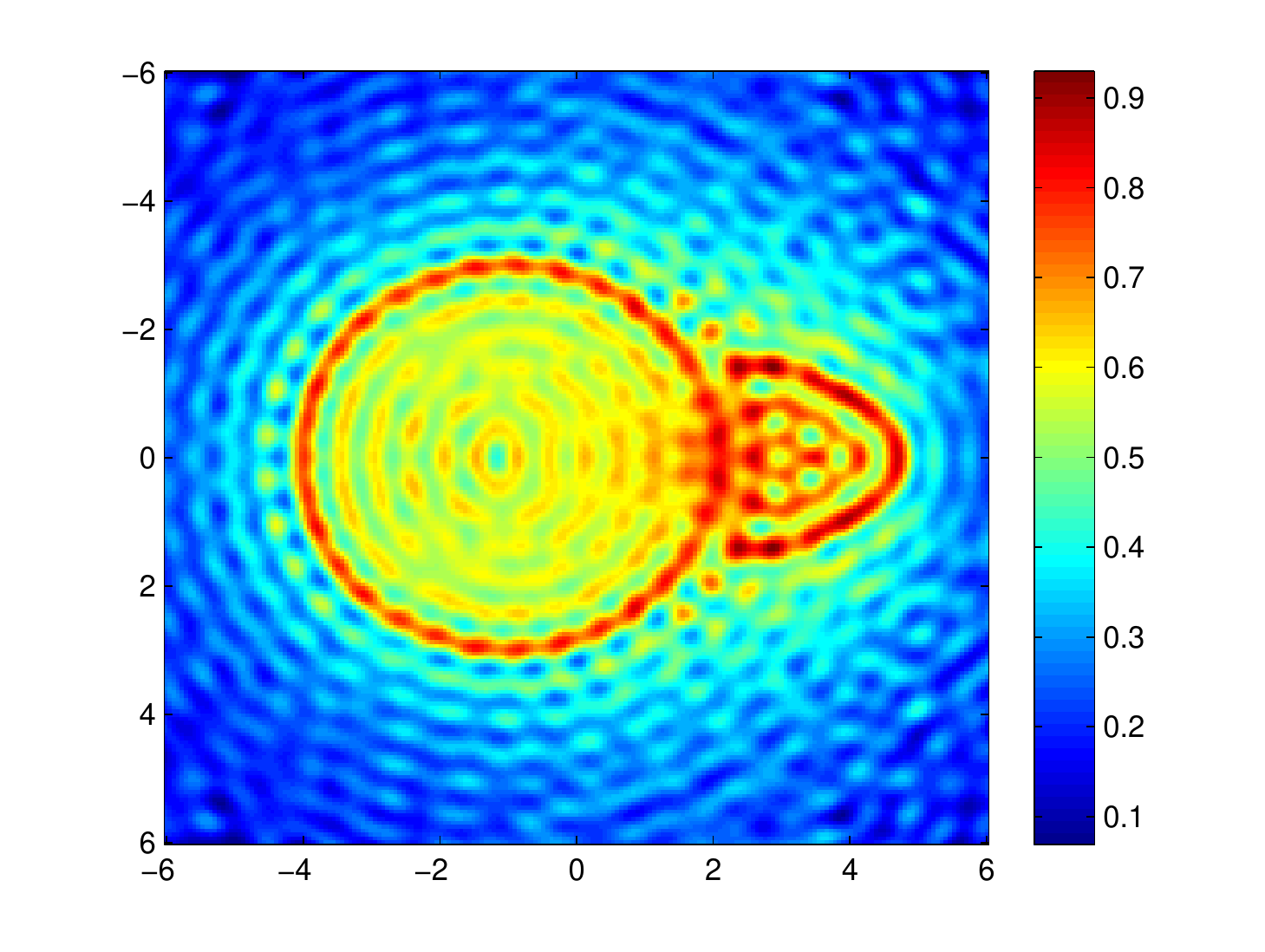}

    \caption{Example 2: The first  and the second picture are imaging results using fixed probe wavelength $\lambda=1$ with reduced number of $(N_s,N_r)=(64,64), (32,128)$, respectively. } \label{figure_67}
\end{figure}

\bigskip
\textbf{Example 2}.
In this example we verify the diffraction limit in our resolution analysis by considering two objects separated with a small fixed distance.
The first model is a circle of radius $\rho=3$ and a kite. The distance between two objects is about 0.5. We use the probe wavelength $\lambda=2,1,0.5$ to image the objects. The search domain is $\Omega=(-6, 6)\times(-6,6)$ with a sampling $201\times 201$ mesh.
The results are shown in Figure \ref{figure_6} from which we observe that with the increase of the probe wave number, the gap between two objects is more and more visible. Figure \ref{figure_67} shows the imaging results when the number of emitters and receivers is reduced.

\begin{figure}
 \centering
    \includegraphics[width=0.24\textwidth, height=1.5in]{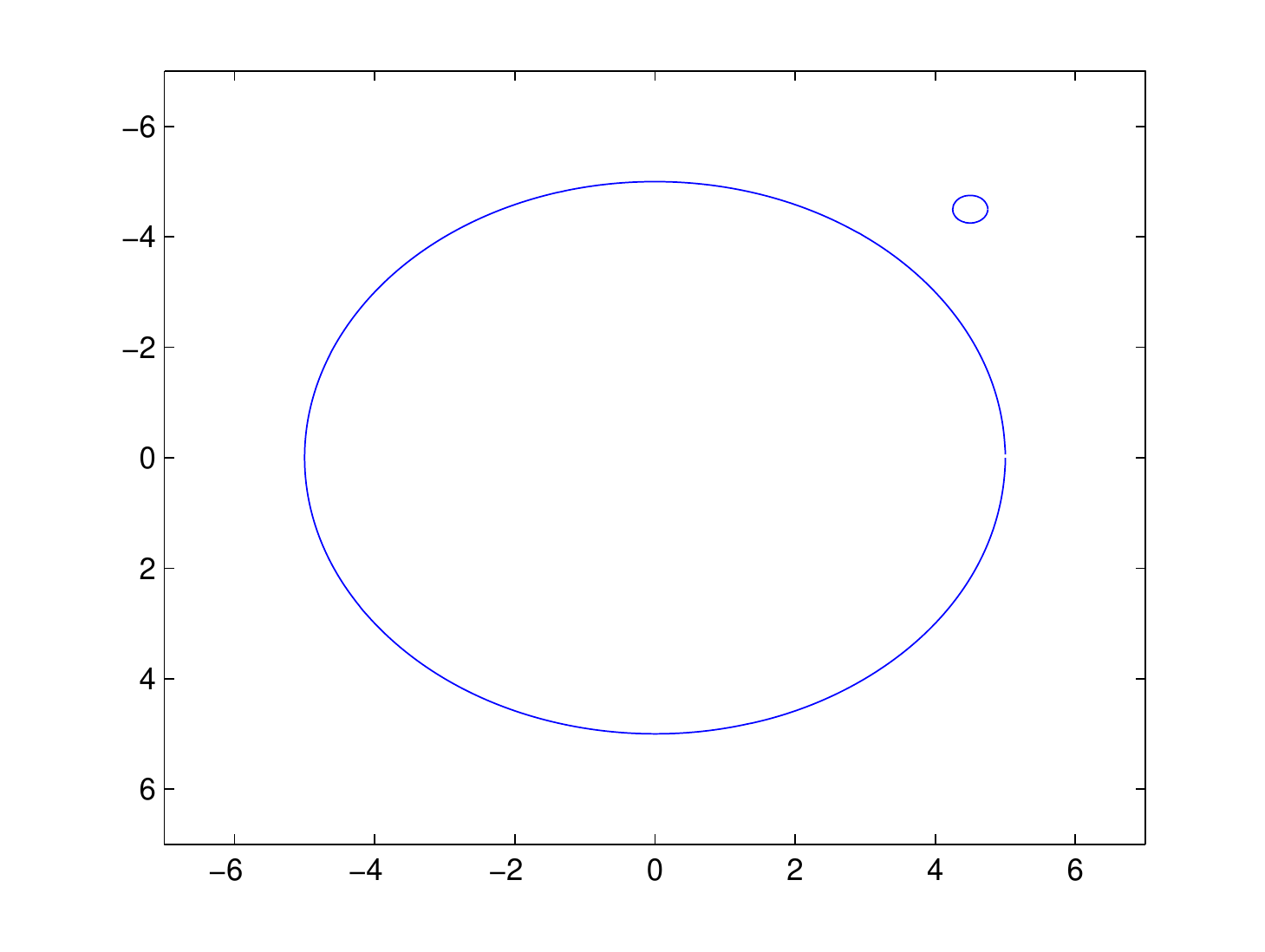}
    \includegraphics[width=0.24\textwidth, height=1.5in]{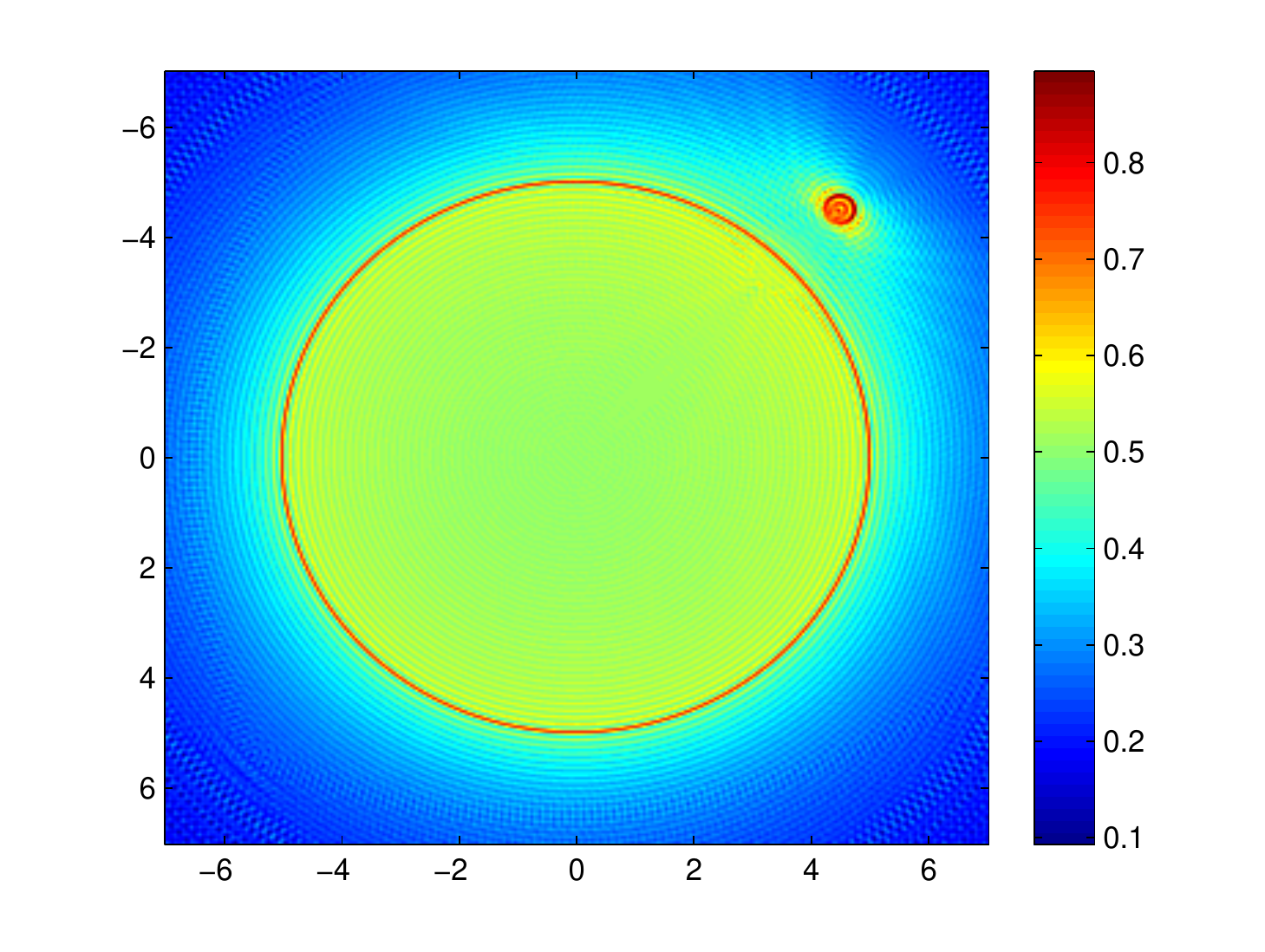}
    \includegraphics[width=0.24\textwidth, height=1.5in]{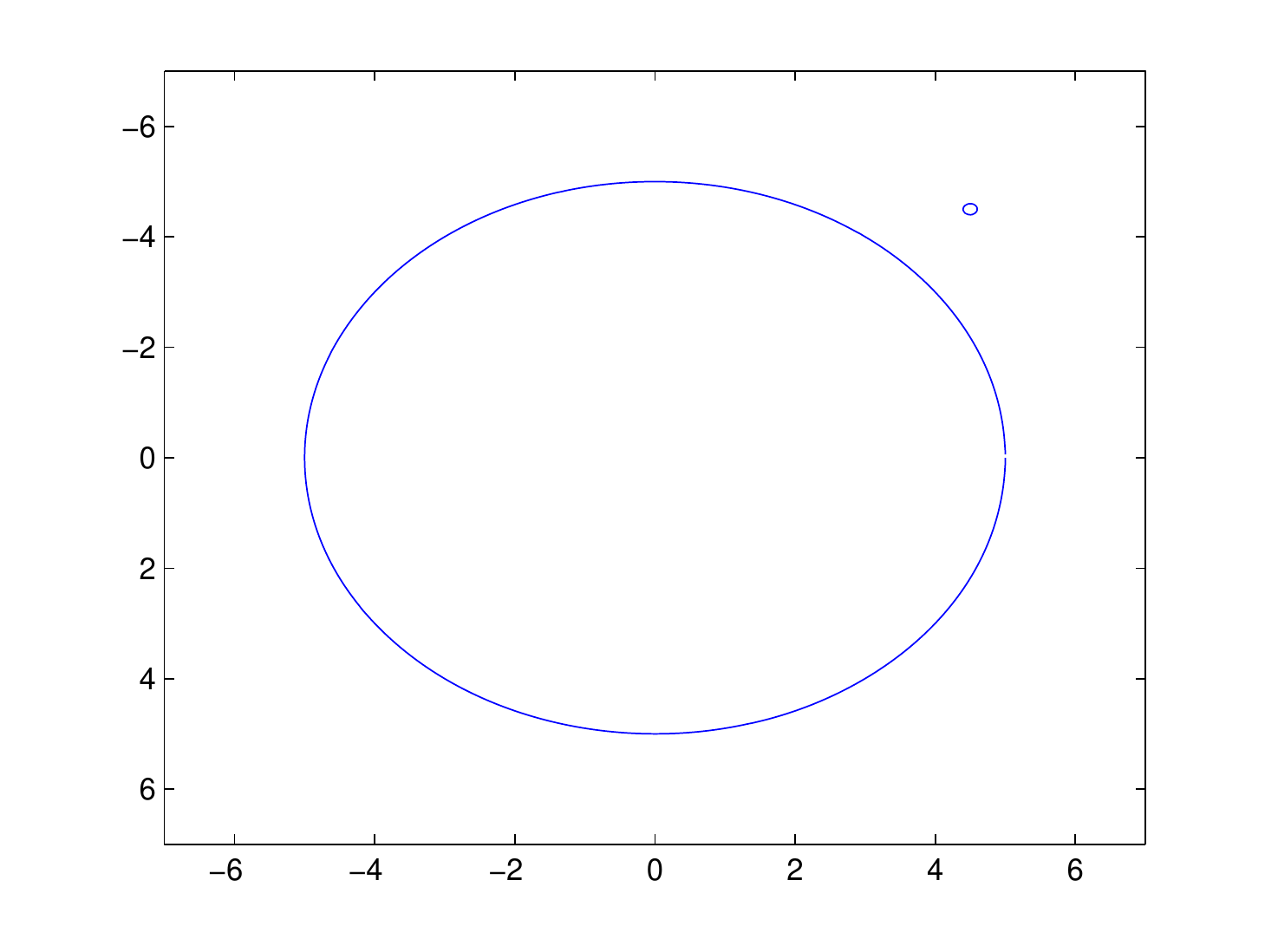}
    \includegraphics[width=0.24\textwidth, height=1.5in]{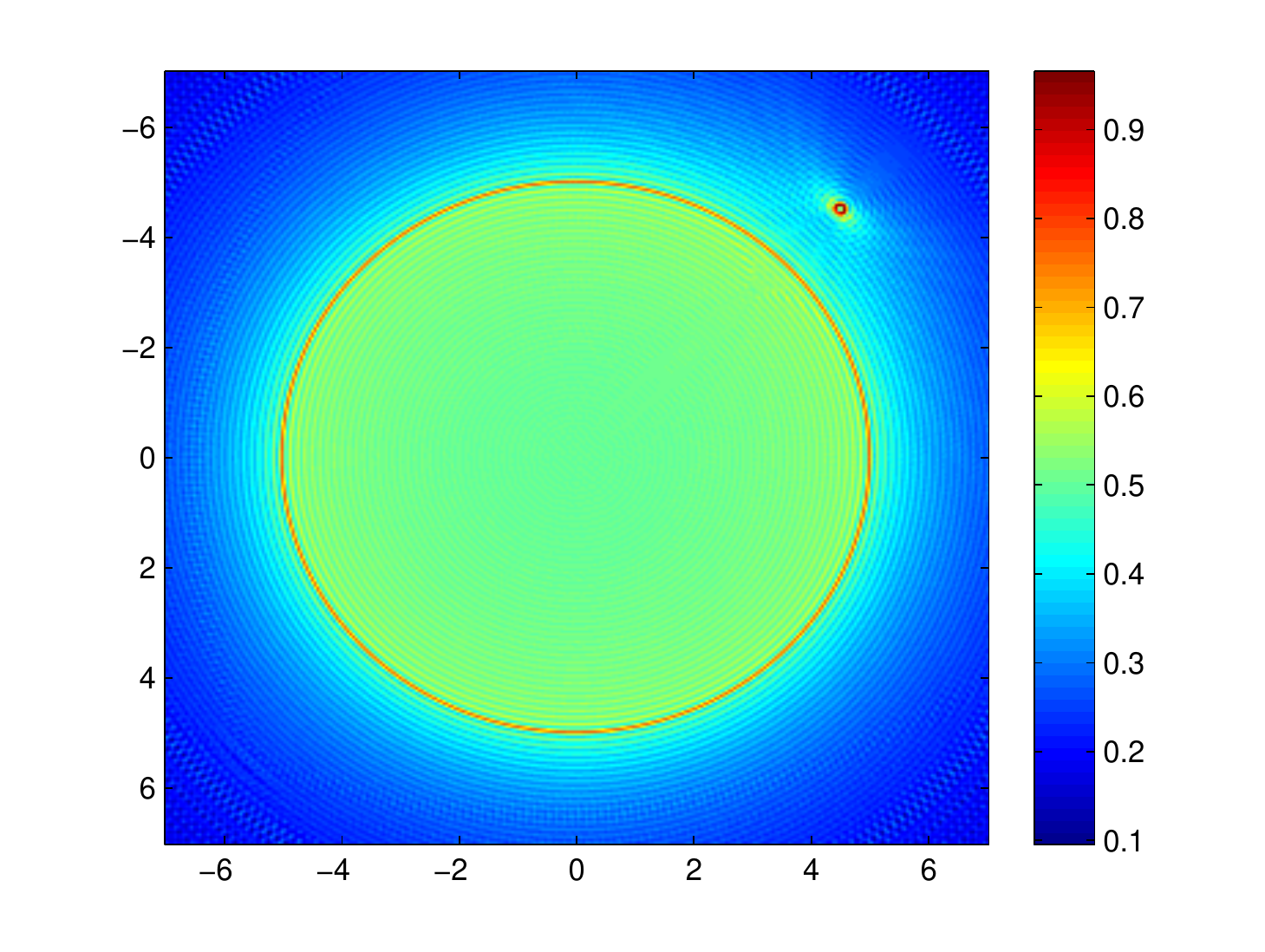}
    \caption{Example 2: The first and third picture show exact obstacles: one big circle with $\rho=5$ and one small circle with radius $\rho=0.25$ (first) or $\rho=0.1$ (third). The second and the fourth picture are imaging results using $\lambda=0.25$ and $N_s=N_r=318, 318$. } \label{figure_7}
\end{figure}

The second model is a big circle of radius $\rho=5$ and a small circle of radius $\rho=0.25$ or $\rho=0.1$. The search domain is $\Omega=(-7, 7)\times(-7,7)$ with a sampling $301\times 301$ mesh. Figure \ref{figure_7} shows the imaging results. We observe that our imaging algorithm clearly locates the boundary of obstacles with different size as long as the high wave number content is available in our received data.

\bigskip
\textbf{Example 3}.
In this example we consider the stability of the imaging functional with respect to the additive Gaussian random noise. We introduce the additive Gaussian noise as follows:
    \begin{equation*}
        u_{noise} = u_s + \nu_{\rm noise},
    \end{equation*}
where $u_s$ is the synthesized data and $\nu_{\rm noise}$ is the Gaussian noise with mean zero and standard deviation $\mu$ times the maximum of  the data $|u_s|$, i.e. $\nu_{\rm noise}  \thicksim \mathcal{N}(0,\mu \max|u_s|)$.

\begin{figure}[h]
 \centering
    \includegraphics[width=0.24\textwidth, height=1.5in]{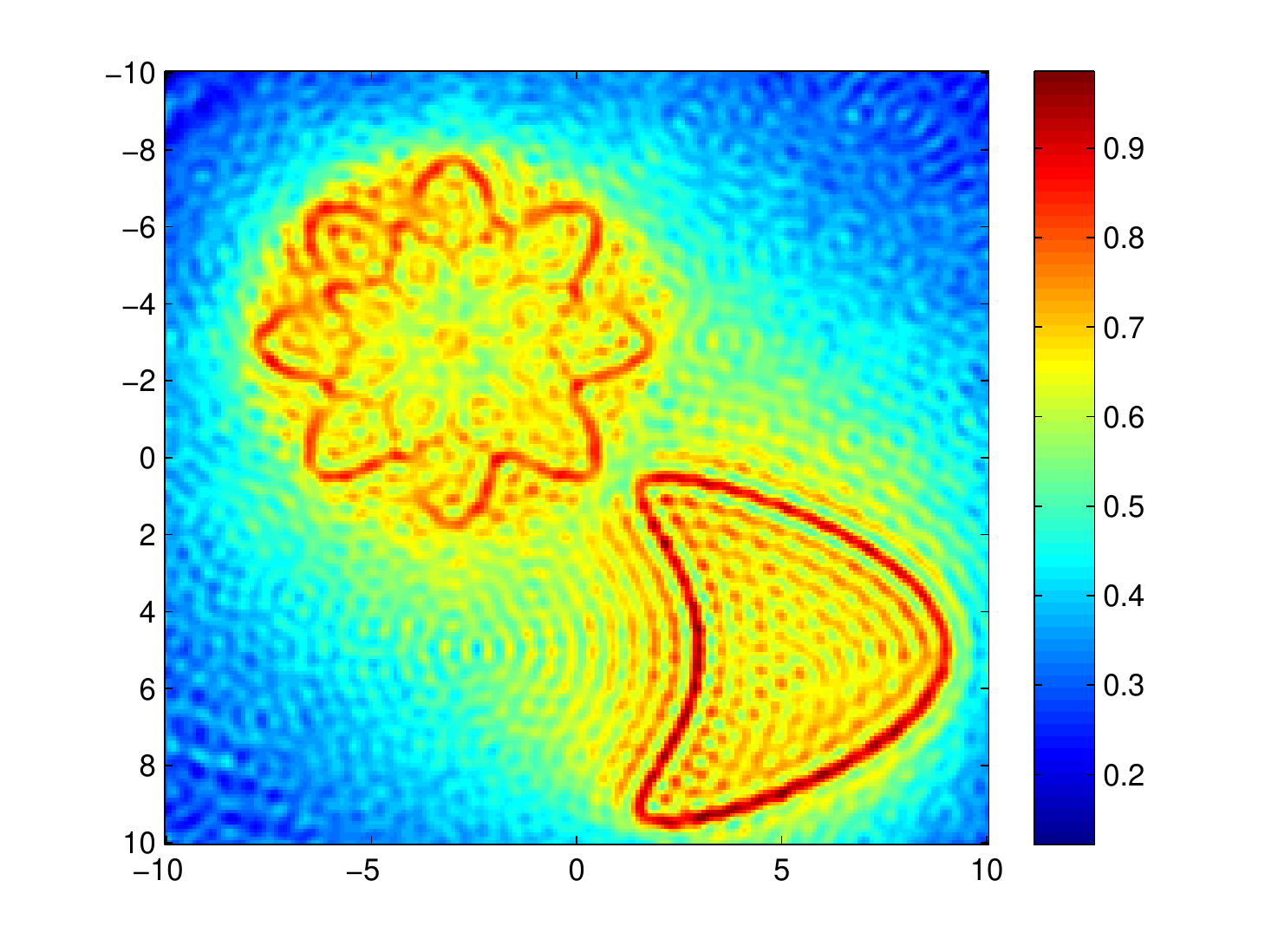}
    \includegraphics[width=0.24\textwidth, height=1.5in]{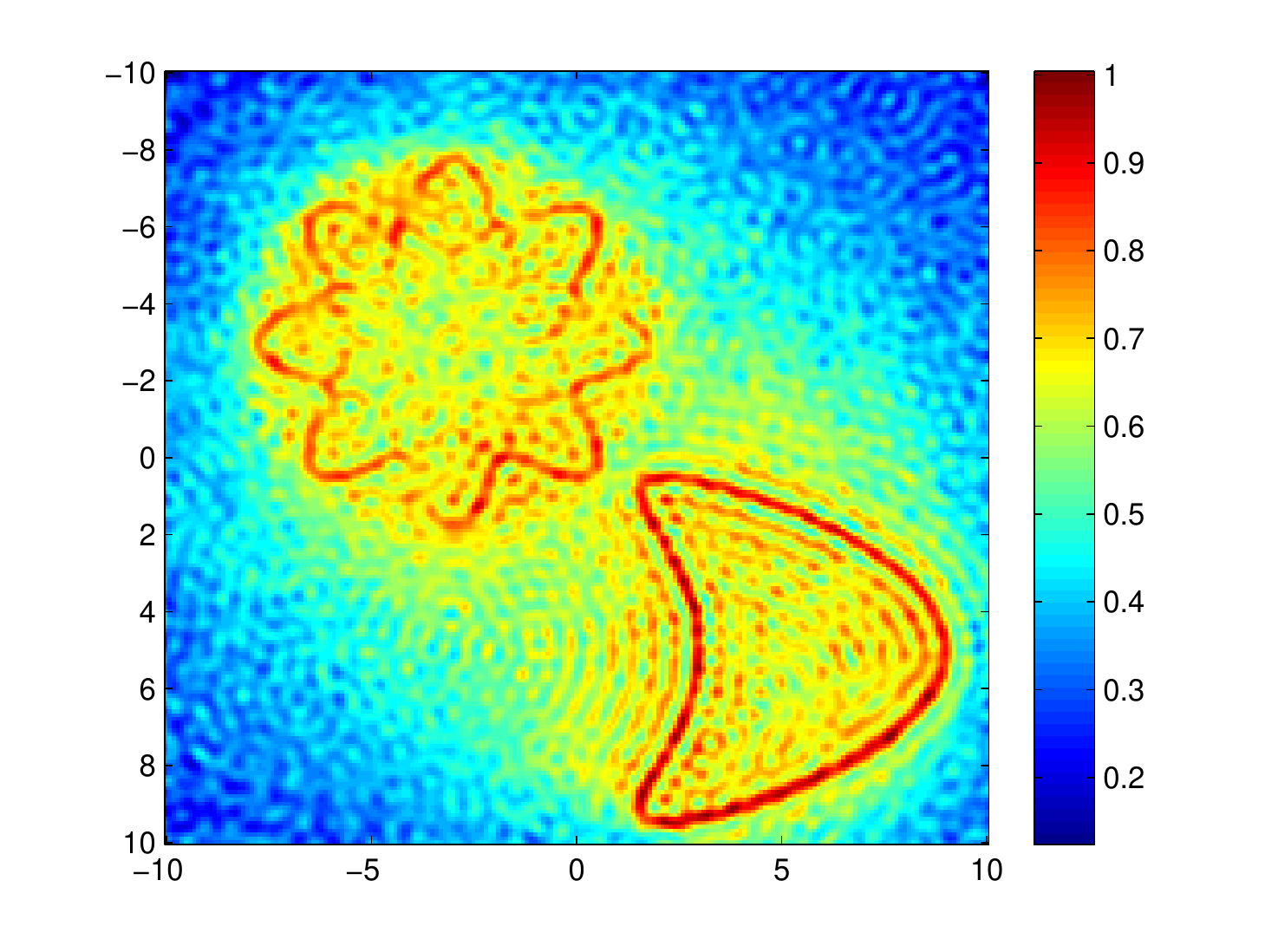}
    \includegraphics[width=0.24\textwidth, height=1.5in]{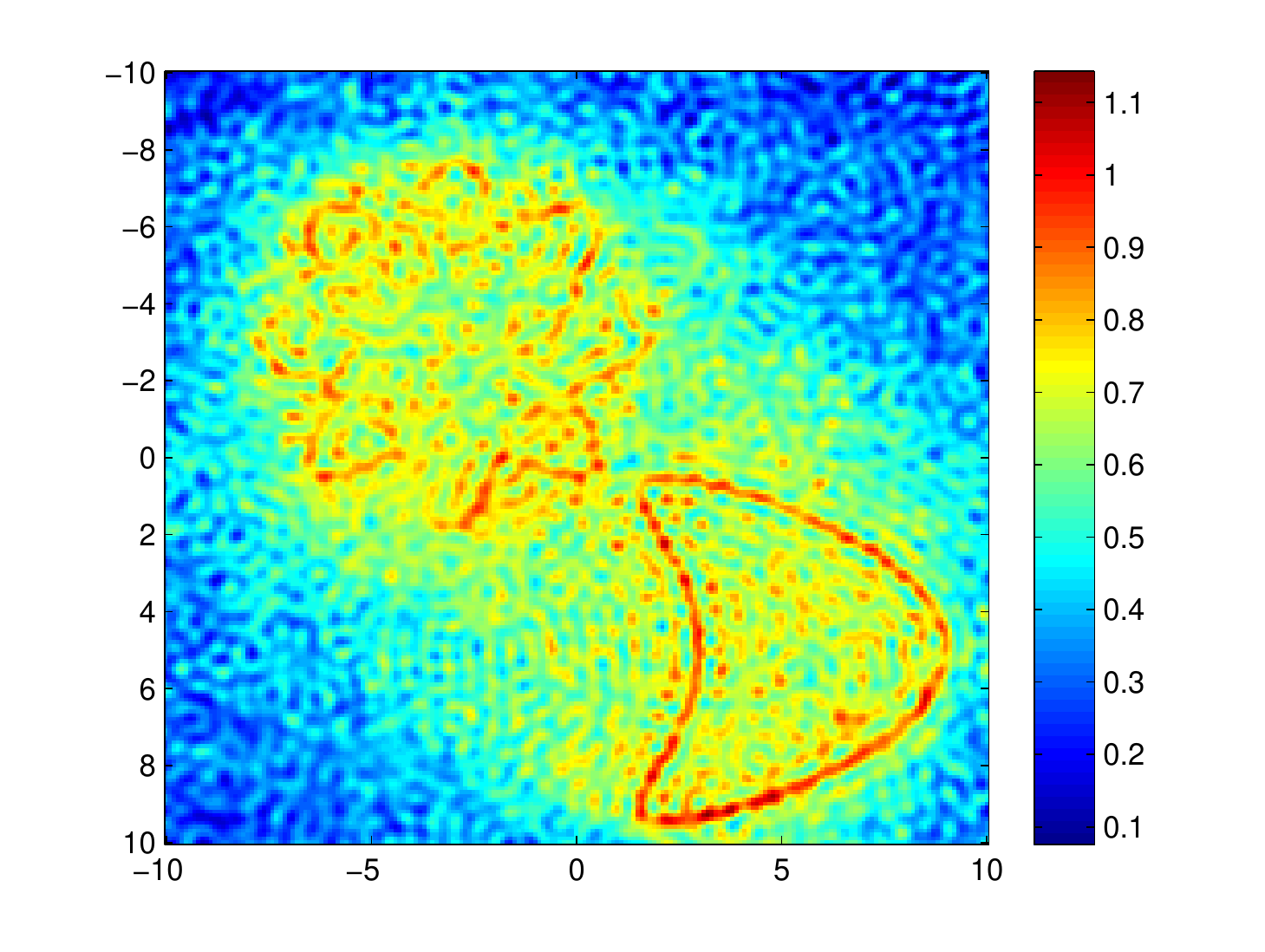}
    \includegraphics[width=0.24\textwidth, height=1.5in]{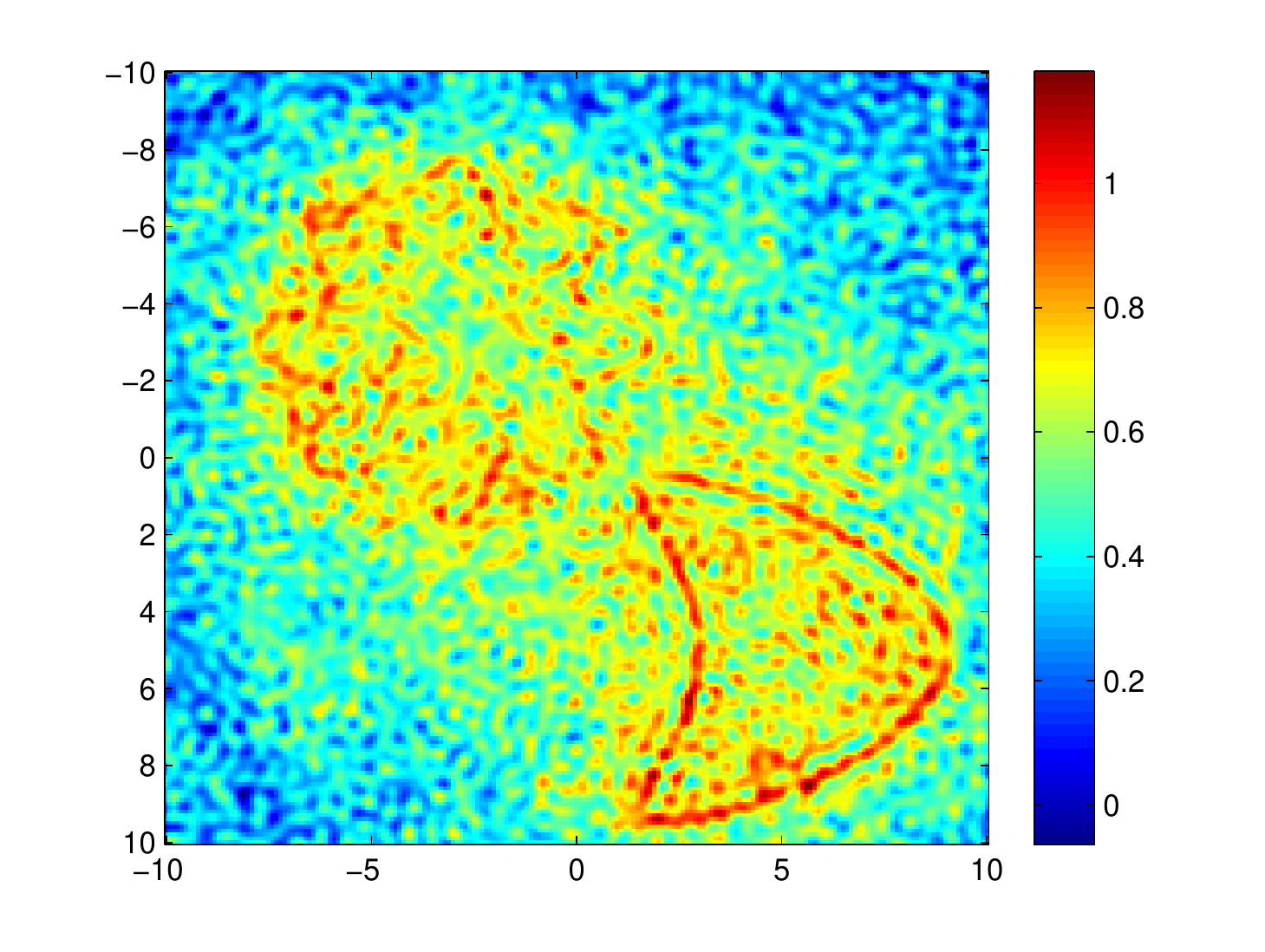}
    \caption{Example 3: The imaging results using data added with additive Gaussian noise and $\mu = 10\%, 20\%, 40\%, 60\%$ from left to right,  respectively. The probe wavelength $\lambda=1$ and $N_s=N_r=128$.} \label{figure_9}
\end{figure}

For the fixed probe wavelength $\lambda=1$, we choose one kite and one 8-leaf in our test.  The search domain is $\Omega=(-10, 10)\times(-10,10)$ with a sampling $201\times 201$ mesh. Figure \ref{figure_9} shows the imaging results with $\mu = 10\%, 20\%,40\%, 60\%$ noise in the single frequency scattered data. The left table in Table \ref{table1} shows the noise level in this case, where $\sigma=\max_{x_r,x_s}|u^s(x_r,x_s)|$, $\|u_s\|_{\ell^2}^2=\frac{1}{N_sN_r}\sum^{N_s,N_r}_{s,r=1}|u^s(x_r,x_s)|^2$, and $\|\nu_{\rm noise}\|_{\ell^2}^2=
\frac{1}{N_sN_r}\sum^{N_s,N_r}_{s,r=1}|\nu_{\rm noise}(x_r,x_s)|^2$.

\bigskip
\begin{table}[h]
\begin{center}
\begin{tabular}{ | c | c | c | c |  }
\hline
$\mu$ & $\sigma$ & $\|u_s\|_{\ell^2}$      & $\|\nu_{\rm noise}\|_{\ell^2}$ \\ \hline
0.1      &   0.003348 &	0.010396	& 0.003898       \\ \hline
0.2       &  0.006697& 0.010396	 & 0.007734  \\ \hline
0.4 &0.013394	& 0.010396 &	0.015386 \\ \hline
0.6 & 0.020091	&0.010396 &	0.02323 \\ \hline
\end{tabular} \ \ \ \
\begin{tabular}{ | c | c | c | c |  }
\hline
$\mu$ & $\sigma$ & $\|u_s\|_{\ell^2}$      & $\|\nu_{\rm noise}\|_{\ell^2}$ \\ \hline
0.1 &	0.003105&0.010452	&0.003589\\ \hline
0.2&	0.006211	&0.010452&	0.007203\\ \hline
0.4&0.012422	&0.010452&	0.014379\\ \hline
0.6&0.018633&0.010452&	0.021451\\ \hline
\end{tabular}
\end{center}
\caption{Example 3: The noise level in the case of single frequency data (left) and multi-frequency data (right).}\label{table1}
\end{table}

\begin{figure}
    \centering
     \includegraphics[width=0.24\textwidth, height=1.5in]{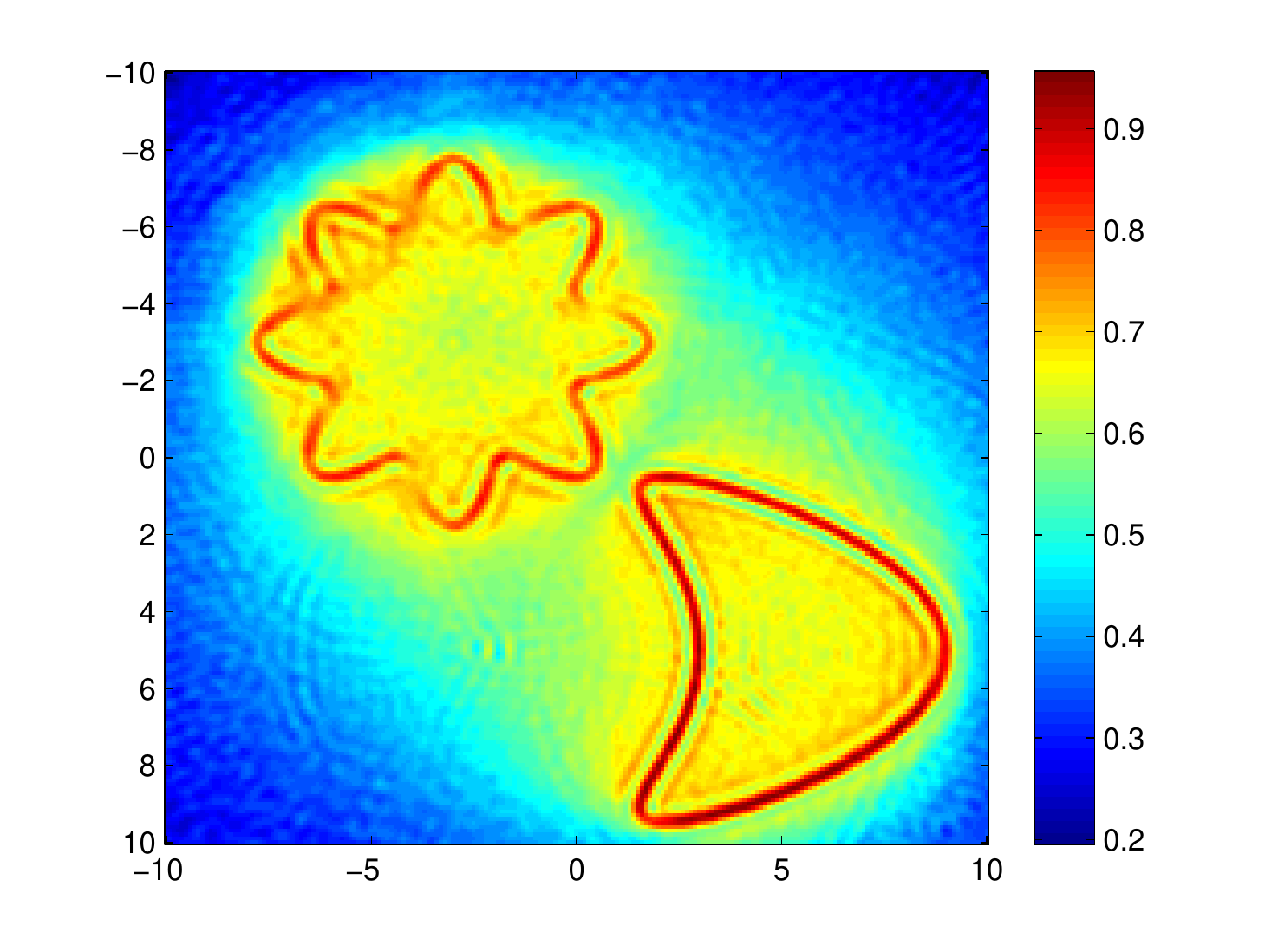}
     \includegraphics[width=0.24\textwidth, height=1.5in]{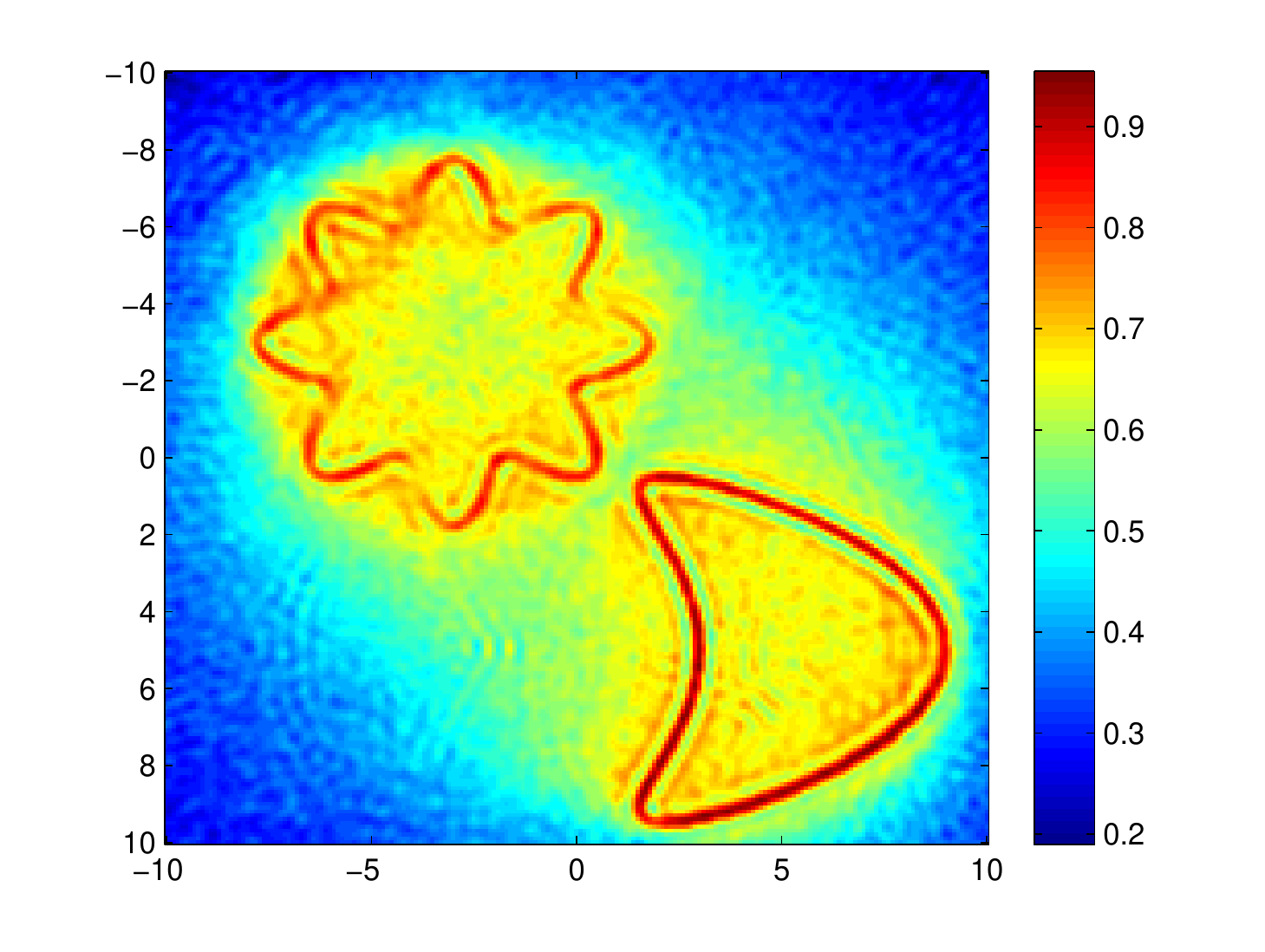}
     \includegraphics[width=0.24\textwidth, height=1.5in]{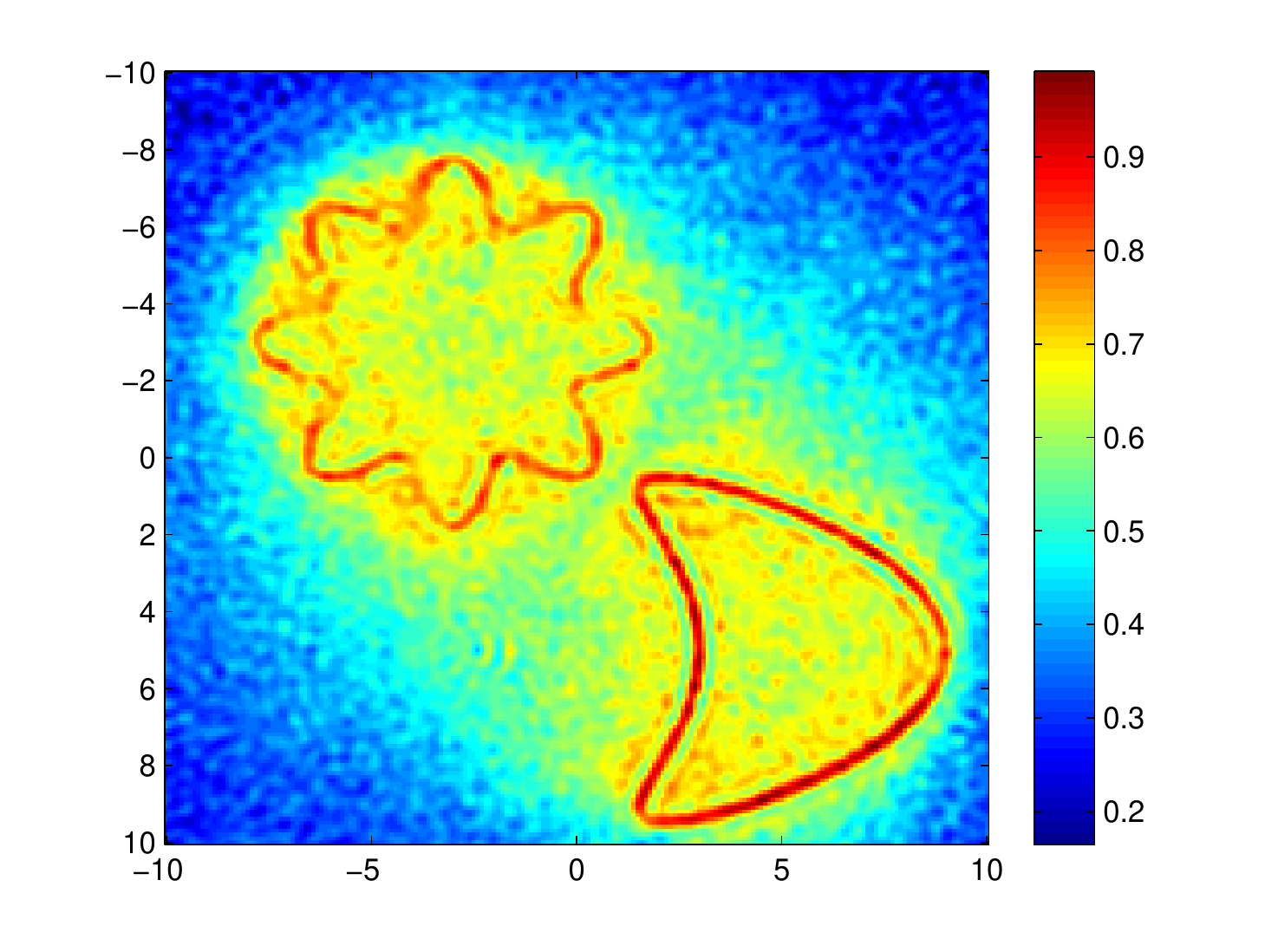}
     \includegraphics[width=0.24\textwidth, height=1.5in]{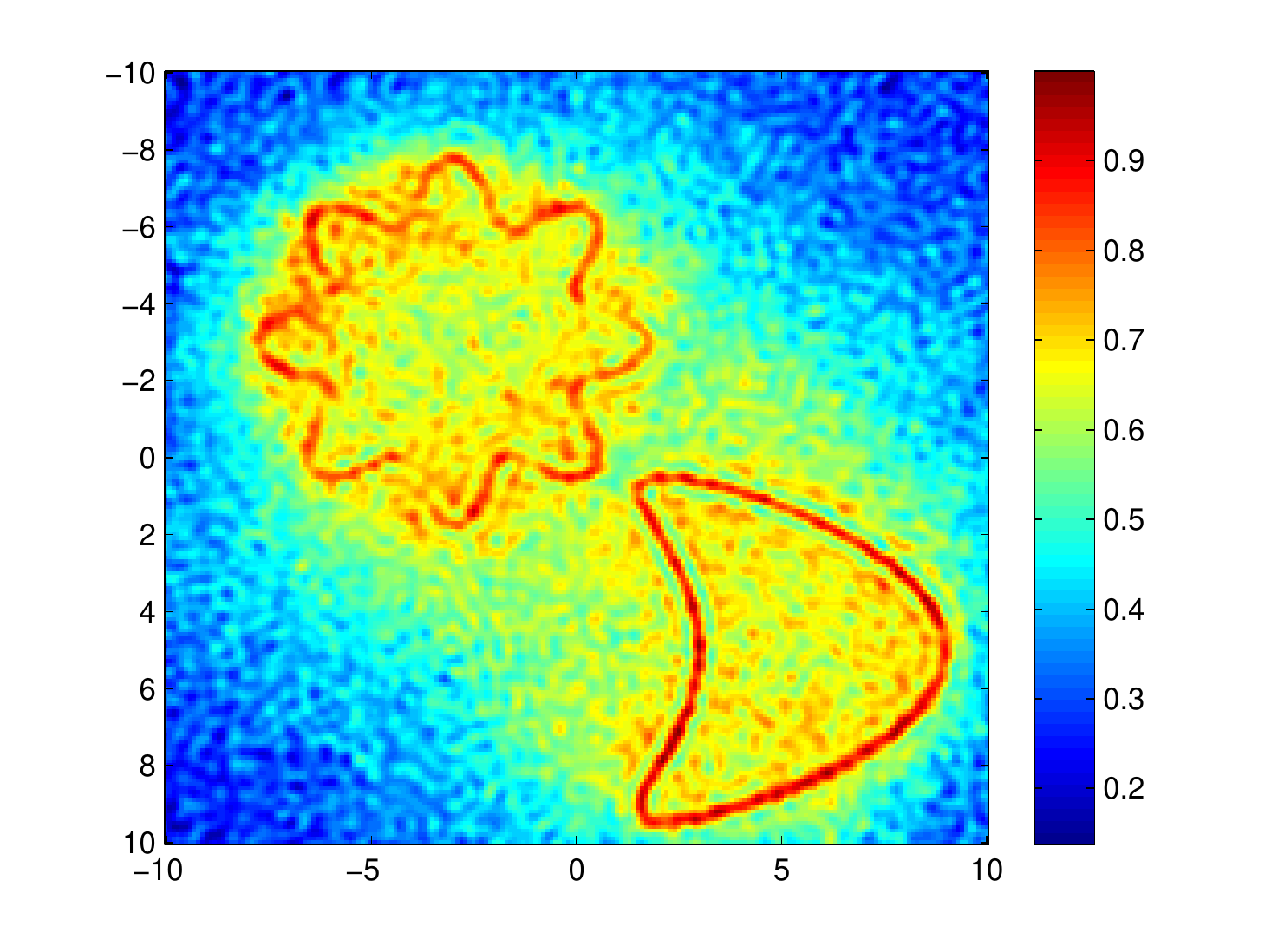}
     \caption{Example 3: The imaging results using multi-frequency data added with additive Gaussian noise and $\mu=10\%, 20\%, 40\%, 60\%$ from left to right, respectively. The probe wavelengths $\lam=1/0.8, 1/0.9, 1/1.0, 1/1.1, 1/1.2$ and
$N_s=N_r=128$.} \label{figure_10}
\end{figure}

The imaging quality can be improved by using multi-frequency data as illustrated in Figure \ref{figure_10} in which we show the imaging results of summing the imaging functionals of probe wavelengths $\lam=1/0.8, 1/0.9, 1/1.0, 1/1.1, 1/1.2$.  The right table in Table \ref{table1} shows the noise level in the case of multi-frequency data, where $\sigma$, $\|u_s\|_{\ell^2}$, and $\|\nu_{\rm noise}\|_{\ell^2}$ are the arithmetic mean of the corresponding values for different frequencies.

\begin{figure}
\centering
    \includegraphics[width=0.3\textwidth, height=1.6in]{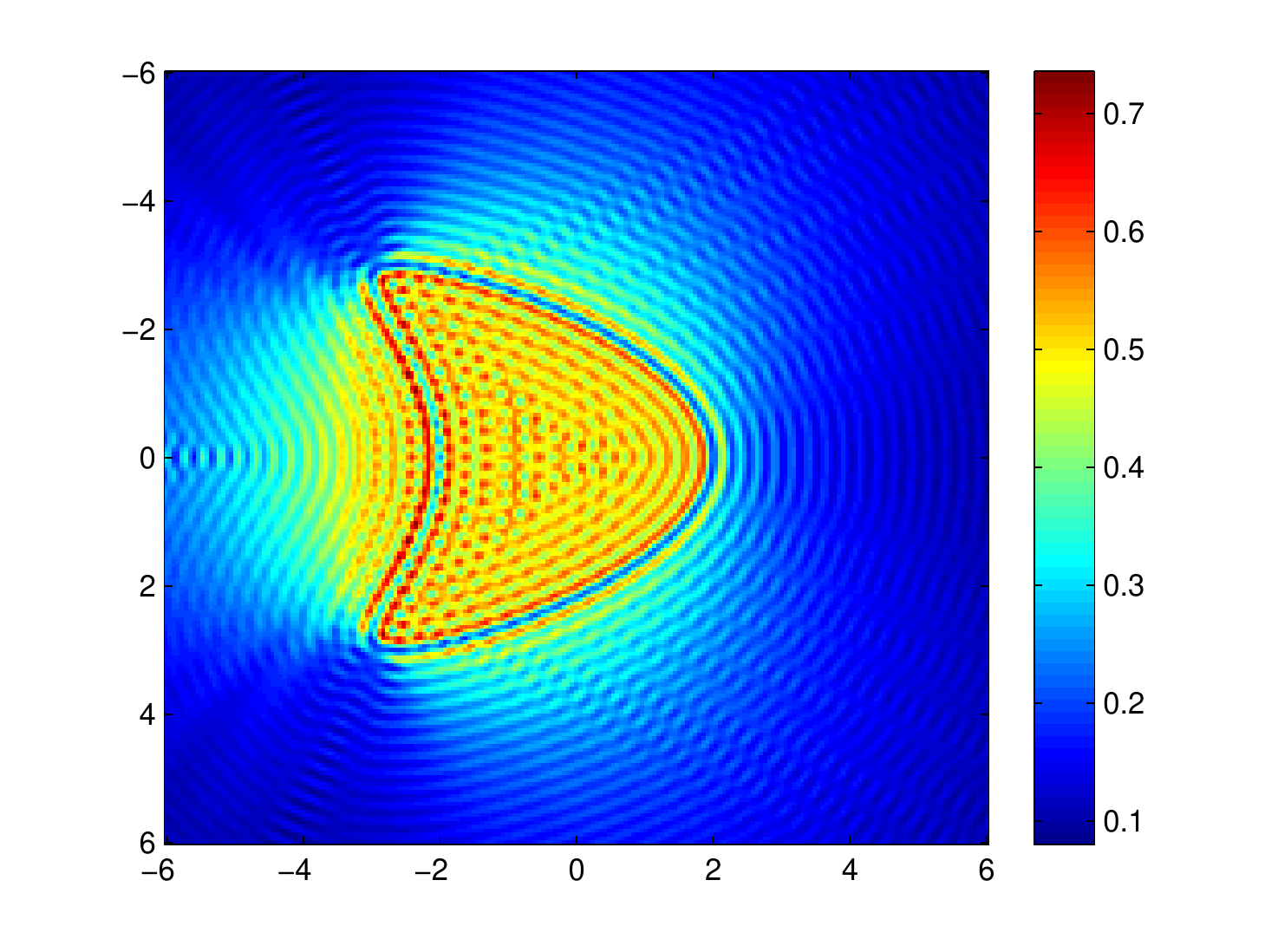}
    \includegraphics[width=0.3\textwidth, height=1.6in]{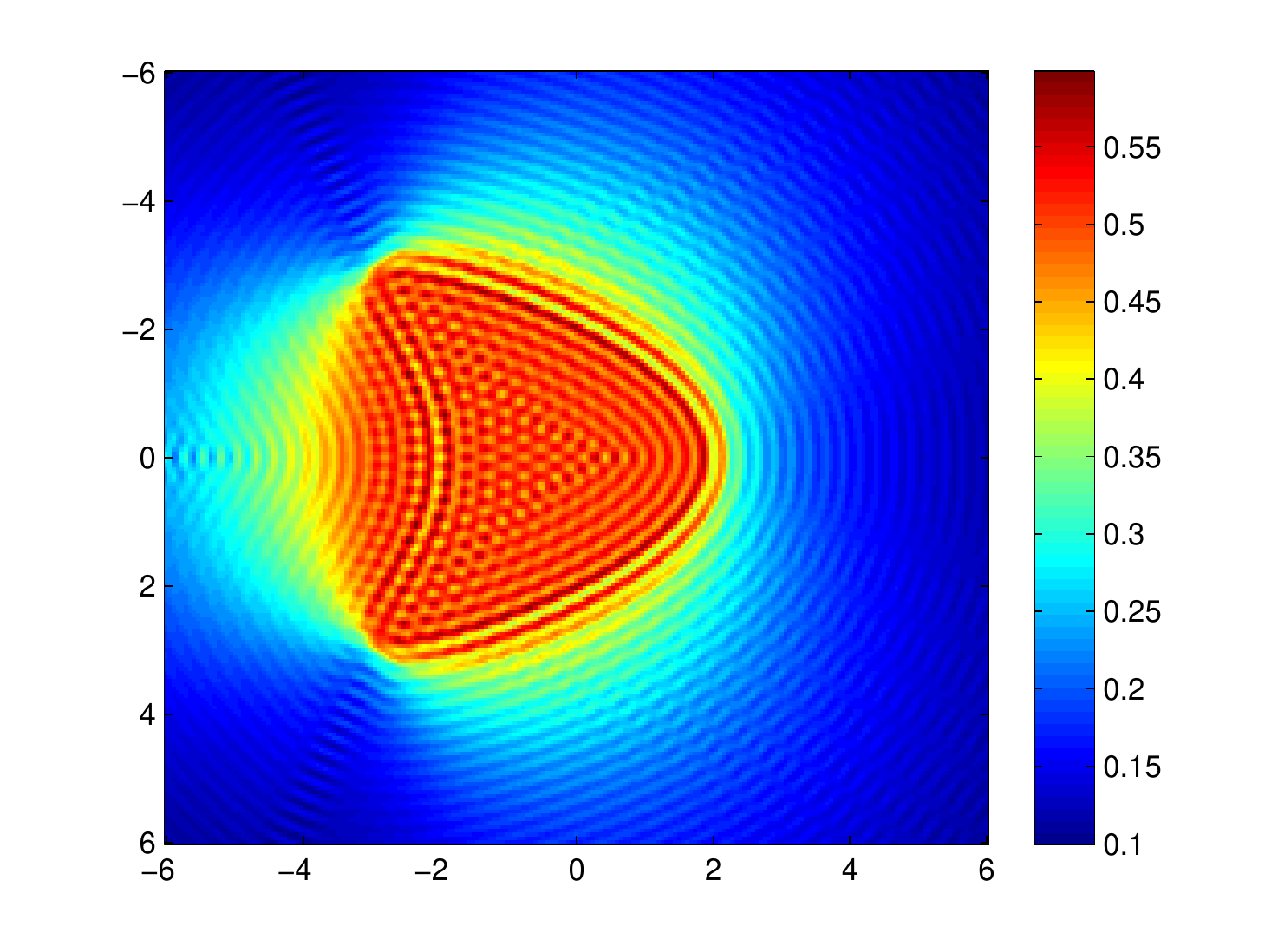}
    \includegraphics[width=0.3\textwidth, height=1.6in]{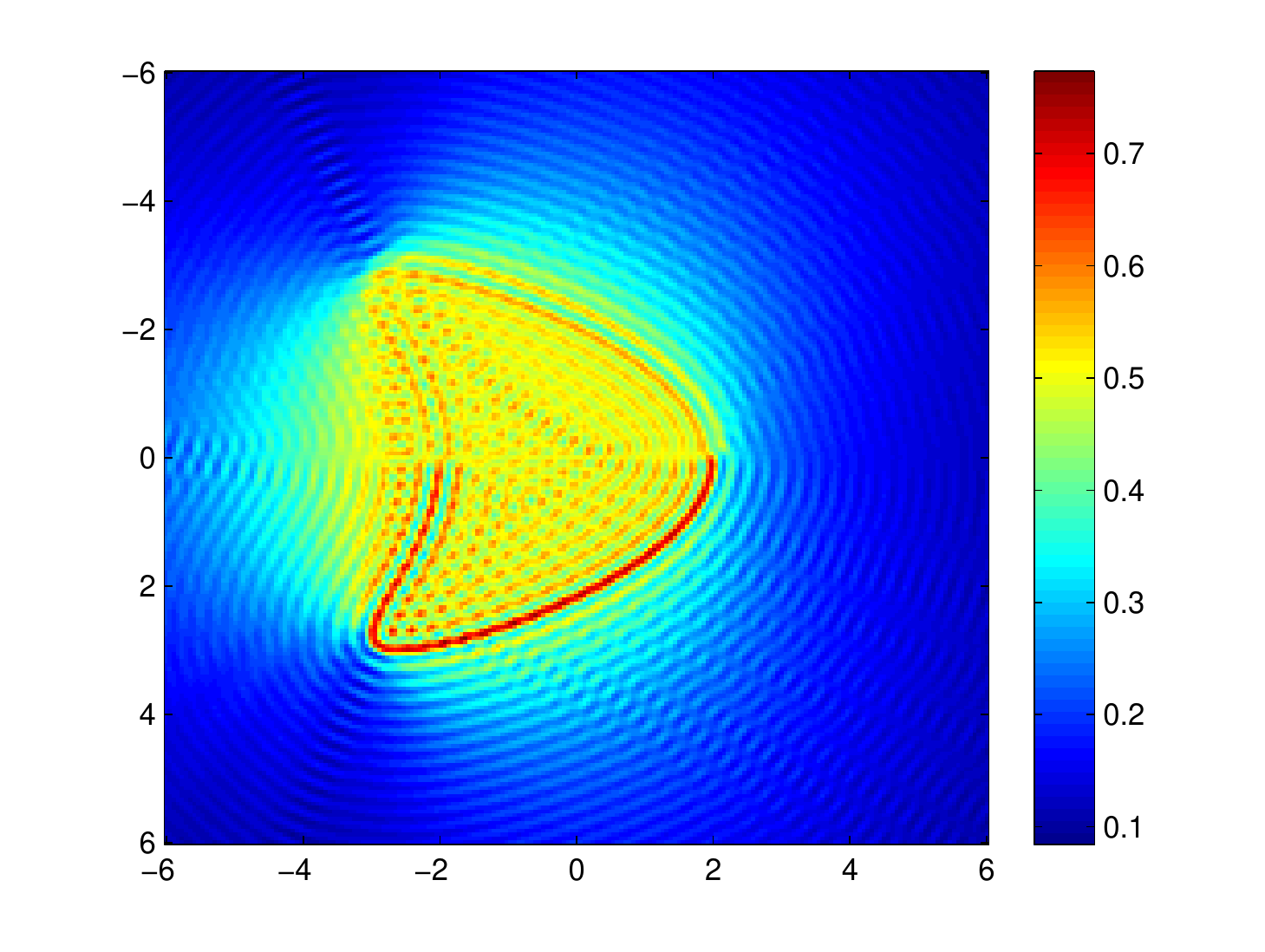}
    \caption{Example 4: The first picture is the Neumann boundary condition $\eta(x)=0$, the second picture is the impedance condition with $\eta(x)=1$, and the third picture is the coated obstacle with the impedance $\eta(x)=1000$ on the upper half part and $\eta(x)=1$ on the lower half part of the boundary of the obstacle. The probe wavelength $\lambda=0.5$ and $N_s=N_r=180$.} \label{figure_11}
\end{figure}

\bigskip
\textbf{Example 4}.
In this example we consider the imaging of non-penetrable obstacles with different impedance conditions. We take the probe wavelength $\lambda=0.5$. The search domain is $\Om=(-6,6)\times (-6,6)$ with a $201\times 201$ sampling mesh. Figure \ref{figure_11} shows the
imaging results which indicate clearly the effectiveness of our imaging algorithm.

\section*{Acknowledgments}
The work of J. Chen is supported in part by China NSF under the grant 11001150, 11171040, and that of Z. Chen is supported in part by National Basic Research Project under the grant 2011CB309700 and China NSF under the grant 11021101. The authors are also grateful to the referees for their constructive comments.

\end{document}